\documentclass[twocolumn]{autart}    

\usepackage{graphicx}          
\usepackage{multicol} 					
\usepackage{color} 							
\usepackage{pgfplots} 					
\usepackage{tikz} 							
\usepackage{amsmath,amssymb,bm} 
\usepackage{graphicx,epstopdf}	
\usepackage{pdfpages}						
\usepackage{subcaption}					
\usepackage{url}
\usepackage{comment}
\usepackage{acronym}
\usepgfplotslibrary{groupplots}
\usepackage{multirow}
\usepackage{algorithm,algpseudocode}
\usepackage{caption}
\usepackage{mathtools}
\usepackage{array}
\usepackage{adjustbox}

\definecolor{mygray}{gray}{0.35}
\newcommand{\red}[1]{\textcolor{red}{#1}} 
\newcommand{\blue}[1]{\textcolor{blue}{#1}} 

\def\IR{\mathbb R}					

\newtheorem{problem}{Problem}

\newcommand{\defeq}{\vcentcolon=}
\newcommand{\defeqRight}{=\vcentcolon}

\newcommand{\I}{\mathbf{I}} 
\newcommand{\mbf}[1]{\mathbf{#1}} 
\newcommand{\bmat}[1]{\begin{bmatrix} #1 \end{bmatrix}} 
\newcommand{\Xcal}{\mathcal{X}}

\newcommand{\A}{\mathbf{A}} 
\newcommand{\B}{\mathbf{B}} 
\newcommand{\C}{\mathbf{C}} 
\newcommand{\D}{\mathbf{D}} 
\newcommand{\bS}{\mathbf{S}} 
\newcommand{\x}{\mathbf{x}} 
\newcommand{\bu}{\mathbf{u}} 
\newcommand{\y}{\mathbf{y}} 
\newcommand{\z}{\mathbf{z}} 
\newcommand{\bP}{\mathbf{P}} 
\newcommand{\Q}{\mathbf{Q}} 
\newcommand{\E}{\mbf{E}} 
\newcommand{\R}{\mbf{R}}
\newcommand{\bH}{\mbf{H}}
\newcommand{\F}{\mbf{F}}

\newcommand{\M}{\mbf{M}}

\newcommand{\W}{\mbf{W}}

\newcommand{\X}{\mbf{X}}
\newcommand{\Gcal}{\mathcal{G}}

\newcommand{\Ell}{\mathcal{L}}

\newcommand{\diag}{\mbox{diag}}

\newcommand{\Tcal}{\mathcal{T}}

\newcommand{\vecdim}[1]{\in \mathbb{R}^{#1}}

\newcommand{\norm}[1]{\left\lVert#1\right\rVert}
\newcommand{\absVal}[1]{\left\lvert#1\right\rvert}
\newcommand{\0}{\mathbf{0}}
\newcommand{\J}{\mathbf{J}}
\newcommand{\Ucal}{\mathcal{U}}
\newcommand{\sumLam}{\sum_{j=0}^n\! \lambda_j}

\newcommand{\intInc}[2]{\mathbb{Z}_{#1}^{#2}}
\newcommand{\intEx}[2]{\mathbb{Z}_{\bar{#1}}^{\underline{#2}}}

\newcommand{\co}{\text{co}}

\newcommand{\map}[2]{\mathbb{R}^{#1} \rightarrow \mathbb{R}^{#2}}

\usepackage{centernot}

\newcommand{\thmref}{Theorem }
\newcommand{\thmrefs}{Theorems }
\newcommand{\defref}{Definition }
\newcommand{\lemref}{Lemma }
\newcommand{\corref}{Corollary }

\newcommand{\asmpref}{Assumption }

\newcommand{\secref}{Section }
\newcommand{\axref}{Appendix }

\newcommand{\tabref}{Table }
\newcommand{\probref}{Problem }

\newcommand{\1}{\mathbf{1}}

\acrodef{dl}[{DL}]{deep learning}
\acrodef{rl}[{RL}]{reinforcement learning}
\acrodef{nn}[{NN}]{neural network}
\acrodef{dnn}[{DNN}]{deep neural network}
\acrodef{tdl}[{TDL}]{temporal difference learning}
\acrodef{pid}[{PID}]{proportional–integral–derivative}
\acrodef{us}[{US}]{Ultrasound}
\acrodef{mse}[{MSE}]{mean squared error}
\acrodef{sgd}[{SGD}]{stochastic gradient descent}
\acrodef{ico}[{ICO}]{iterative convex overbounding}
\acrodef{mi}[{MI}]{matrix inequality}
\acrodef{lmi}[{LMI}]{linear matrix inequality}
\acrodef{mjls}[{MJLS}]{Markov jump linear system}
\acrodef{io}[{IO}]{input-output}
\acrodef{iqc}[{IQC}]{integral quadratic constraints}
\acrodef{cnn}[{CNN}]{convolutional neural network}
\acrodef{il}[{IL}]{Imitation learning}
\acrodef{mpc}[{MPC}]{model predictive control}
\acrodef{sdp}[{SDP}]{semi-definite programming}
\acrodef{relu}[{ReLU}]{rectified linear unit}
\acrodef{us}[US]{ultrasound}
\acrodef{mdp}[MDP]{Markov Decision Process}
\acrodef{iid}[iid]{identical and independently distributed random variable}
\acrodef{pid}[PID]{Proportional Integral Derivative}

\acrodef{lqr}[LQR]{linear-quadratic regulator}

\acrodef{cpa}[CPA]{continuous piecewise affine}
\acrodef{hji}[HJI]{Hamilton Jacobi Inequalitie}

\acrodef{vndt}[VNDT]{Vidyasagar's Network Dissipativity Theorem}
\acrodef{pwa}[PWA]{piecewise affine}
\acrodef{kyp}[KYP]{Kalman-Yakubovich-Popov}
\acrodef{sos}[SOS]{sum of squares}
\acrodef{ico}[ICO]{iterative convex overbounding}
\acrodef{dp}[DP]{dynamic programming}

\newcounter{tableeqn}[table]

\newcounter{tablesubeqn}[tableeqn]

 \pgfplotsset{compat=1.18}

\begin{document}

\begin{frontmatter}

\title{Local Dissipativity Analysis of Nonlinear Systems \thanksref{footnoteinfo}} 

\thanks[footnoteinfo]{ This work was supported by the ONR YIP Grant No. N00014-23-1-2043 and NSF
Grant No. 2303158.}

\author[Duke]{Amy K. Strong}\ead{aks121@duke.edu} and    
\author[Duke]{Leila J. Bridgeman}\ead{leila.bridgeman@duke.edu}               

\address[Duke]{Department of Mechanical Engineering and Materials Science at Duke University, Durham, NC, 27708, USA}  
          

\begin{abstract}                          
Dissipativity is an \ac{io} characterization of nonlinear systems that enables compositional robust control through \ac{vndt}. However, determining the dissipativity of a system is an involved and, often, model-specific process. We present a general method to determine the local dissipativity properties of smooth, nonlinear, control affine systems. We simultaneously search for the optimal \ac{io} characterization of a system and synthesize a \ac{cpa} storage function via a convex optimization problem. To do so, we reformulate the dissipation inequality as a \ac{mi} and develop novel \ac{lmi} bounds for a triangulation to impose the dissipativity conditions on the \ac{cpa} storage function Further, we develop a method to synthesize a combined quadratic and \ac{cpa} storage function to expand the systems the optimization problem is applicable to. Finally, we establish that our method will always find a feasible \ac{io} characterization and storage function given that the system is sufficiently strictly locally dissipative and demonstrate the efficacy of our method in determining the conic bounds and gain of various nonlinear systems.
\end{abstract}

\end{frontmatter}

\section{Introduction}
\Acf{io} stability theory is an important tool in robust control. It uses a compositional approach to stability analysis -- that is, the closed loop \ac{io} stability of a network depends on the coarse, open loop \ac{io} characterizations of each individual subsystem \cite{moylan1978stability,vidyasagar1981input}. This enables flexibility when addressing different types of subsystems -- linear, nonlinear, time-varying, etc -- with the additional benefit that \ac{io} characterizations can be selected that are robust to various sources of uncertainty. But therein lies the challenge: how do we find a valid \ac{io} characterization?
For nonlinear systems, determining the \ac{io} characterization is an involved and, often, model specific process. For nonlinear, control affine systems, this work provides a comprehensive method to determine one such \ac{io} characterization: local QSR dissipativity, which encompasses local versions of a range of  well-known \ac{io} characterizations, including passivity \cite{brogliato2007dissipative}, $\Ell_2$ gain \cite{van19922}, conic sectors \cite{zames1966input}, and more \cite{vidyasagar1977l2}.

For linear systems, we can establish \ac{io} properties through frequency-domain analysis \cite{brogliato2007dissipative}, solving \acp{lmi} that arise from \acp{hji} \cite{Gupta1996}, or data-driven analysis \cite{koch2021provably}. 
While \ac{io} theory takes an operator theoretic approach to dynamical systems analysis \cite{zames1966input,vidyasagar1981input,moylan1978stability}, the alternative state-space perspective on QSR dissipativity from \cite{willems1972dissipative,hill1976stability} provides an avenue for analysis by linking a systems internal and supplied energy through a so-called storage function. For linear systems, this relationship led to the celebrated \ac{kyp} Lemma \cite{brogliato2007dissipative}, which provides necessary and sufficient conditions for dissipativity. The \ac{kyp} Lemma equates dissipativity with the existence of a positive, semi-definite, quadratic storage function, which can be found by solving \acp{lmi} via optimization. Analogous to linear systems, dissipativity of a nonlinear, control affine system is equivalent to the existence of a positive semi-definite storage function \cite{hill1976stability,hill1980connections}. However, because the storage function lacks a set functional form, there is no systematic way to find it, so model-dependent ad-hoc ingenuity is needed. As a result, much effort has been focused on developing model agnostic analysis methods for nonlinear systems.
\ac{io} analysis of nonlinear systems requires synthesis of the system's storage function that adheres to the associated \ac{hji}. The problem of \ac{io} analysis of nonlinear systems can be approached from an optimal control perspective -- using \ac{dp} schemes to approximate a system's storage function \cite{james1995numerical,dower2008dynamic}. However, \cite{dower2008dynamic} is limited to determining gain over a finite horizon, while \cite{james1995numerical} uses iterative methods that require a good initialization and do not always have convergence guarantees. Another approach is to transform the nonlinear system into an easier form for analysis \cite{sutavani2023small,zahedzadeh2009upper}. For example, in \cite{sutavani2023small}, nonlinear systems are represented using Koopman operators so that analysis tools for linear systems can be used \cite{sutavani2023small}; however, this method is only applicable in a limited region of the state space where the transformation is valid. On the other hand, transformations in \cite{zahedzadeh2009upper} require specific classes of nonlinear systems.

Another common approach in \ac{io} analysis is to parameterize the storage function using a a polynomial function \cite{summers2013quantitative,tan2006local,papachristodoulou2005tutorial} and then learn the function parameters that adhere to the \ac{hji} via  \ac{sos} programming. While flexible, this method only applies to polynomial systems \cite{summers2013quantitative,tan2006local}. Further, it is limited to verification of \ac{io} properties \cite{papachristodoulou2005tutorial} or uses non-convex optimization programs to determine \ac{io} properties \cite{summers2013quantitative}, which may result in overly conservative system characterizations. \Ac{sos} methods have since been extended to the data driven dissipativity characterizations \cite{martin2023data}, but do not give deterministic guarantees and require data on a system's first and second derivative, which is often noisy.

Preliminary work tackled the special case of identifying gain of nonlinear systems by formulating the search for a \ac{cpa} storage function as convex optimization problems \cite{lavaei2022simultaneous,strong2024improved}. Constraints were imposed on the storage function at each vertex of the triangulation to ensure certain properties, leveraging a similar process as in previous work on Lyapunov function synthesis \cite{giesl2012construction,giesl2014revised}. We aim to expand upon this approach, admitting nonzero control affine terms where \cite{lavaei2022simultaneous} did not, and examining QSR dissipativity instead of gain.

\subsection*{Contributions}

This paper determines the local QSR dissipativity characterization of nonlinear, control affine systems via synthesis of a \ac{cpa} storage function that adheres to the dissipation condition over a region of the state space. This analysis is accomplished via a convex optimization problem that simultaneously searches for the QSR parameters and \ac{cpa} storage function that satisfies that the dissipation condition --  ensuring that the inequality holds for the entirety of the region. In essence, our approach triangulates a region of the state space, and then upper bounds the nonlinear dynamics of the system on each simplex with a linear system. We are then able to synthesize a storage function across the linearized systems.

In \secref \ref{sec:ErrorBounds}, we develop a novel \ac{lmi} error bound to impose \ac{mi} constraints on a \ac{cpa} function on a triangulation. We also develop a method to enforce inequalities on a closed norm ball about the origin, which allows us to consider quadratic storage functions. The development of these error bounds builds on work in \cite{strong2024improved}.  In \secref 
\ref{sec:dissAnalysis}, we develop conditions for a \ac{cpa} storage function to adhere to the dissipation condition. Because of the \ac{lmi} error bounds, we are able to reformulate the dissipation condition as an \ac{mi}-- convexifying the previous optimization problem. Optimization problems are formulated to search for general dissipativity characterizations of nonlinear systems. In \secref \ref{sec:solutionGuarantess} theoretical results are developed to confirm that if a nonlinear control affine system is sufficiently strictly locally dissipative, then this optimization problem can be used to verify dissipativity characteristics. Finally, numerical results are given in \secref \ref{sec:numRes}.

\subsection*{Notation}

The interior, boundary, and closure of the set $\Omega\subset\mathbb{R}^n$ are denoted as $\Omega^\circ$, $\delta\Omega$, and $\bar{\Omega}$. The symbol $\mathfrak{R}^n$ denotes the set of all compact subsets $\Omega\subset\mathbb{R}^n$ satisfying i) $\Omega^\circ$ is connected and contains the origin and ii) $\Omega = \bar{\Omega}^\circ.$ Let $\mathbb{R}_+$ represent the set of non-negative real numbers. The notation $\intInc{a}{b}$ ($\intEx{a}{b}$) denote the set of integers between $a$ and $b$ inclusive (exclusive).  The p-norm of a vector $x\vecdim{n}$ is denoted as $\norm{\cdot}_p,$ where $p\in\intInc{1}{\infty}.$ 
Let $B_{\epsilon}(0)$ denote a closed two-norm ball about the origin with radius $\epsilon.$

The positive definiteness (semi-definiteness) of a matrix is denoted by $\bP \succ 0$ ($\bP \succeq 0$). Identity and zero matrices are denoted as $\I$ and $\0.$ Let $1_n$ denote a vector of ones in $\mathbb{R}^n,$ and let $\text{He}(\X) = \X + \X^\top.$ Recall that for a matrix operator norm, $\norm{\A\B} \leq \norm{\A}\norm{\B}$ \cite{petersen2008matrix}.  From Lemma 2.3 of \cite{giesl2012construction}, if $\bar{h}$ is the maximum absolute value of any element within the matrix $\bH \vecdim{n\times m}$, then $\norm{\bH}_2 \leq \sqrt{nm}\bar{h}.$ 

Let the $\Ell_2^n$ space be a set containing all measurable functions $\bu:\mathbb{R}_+ \rightarrow \mathbb{R}^n$ that satisfy  $\int_0^\infty  \bu(t)^\top \bu(t)  dt < \infty$. The extended $\Ell^n_{2e}$ space is defined as the set of all measurable functions $\bu:\mathbb{R}_+ \rightarrow \mathbb{R}^n$ for which the truncation to $t\in[0,T]$ is in $\Ell_2^n$ for all $T\geq 0$.

For the function $f:\map{n}{n},$ let $f^{(k)}:\map{n}{}$ denote the $k^{th}$ element of $f,$ and let $\bH_{f^{(k)}}$ denote the Hessian for this element, where $k \in \mathbb{Z}_1^n.$ Further, let $f{\in}\mathcal{C}^r(\Xcal)$ denote a real valued function is $r$-times differentiable over the domain $\Xcal$. The right-hand (left-hand) upper Dini-derivatives for a function, $f$, is defined $D^{+}f(\y) \defeq \lim \sup_{k \rightarrow 0^{+}} \frac{f(x {+} kg(x)) {-} f(x)}{k}$ \\ $ \left(\lim \sup_{k \rightarrow 0^{-}} \frac{f(x {+} kg(x)) {-} f(x)}{k}\right),$ where $k {\in} \mathbb{R}$ and $\dot{x} {= }g(x)$ \cite{giesl2014revised}.

\section{Problem Statement and Background}\label{sec:probAndBackground}

This paper aims to characterize the local dissipativity properties of the mapping $\Gcal(x_0):\Ell_{2e}^m \rightarrow \Ell_{2e}^p$, where $\y = \Gcal(x_0)\bu$ is defined by $x(0)=x_0 \in \mathbb{R}^n$ and
\begin{flalign}\label{eq:dynSys}
    \Gcal(x_0): \begin{cases}
        \dot{x} & = f(x) + \bar{G}(x)u,  \\
        y & = h(x) + \bar{J}(x)u.
    \end{cases}
\end{flalign}
Here, $\bar{G}(x) = \B + G(x)$ and $\bar{J}(x) = \D + J(x).$ Let $f(0) = 0,$ $G(0) = 0,$ $h(0) = 0,$ and $J(0) = 0$.

We will establish local dissipativity by developing constraints on \ac{cpa} storage functions implying QSR dissipativity. 
Local dissipativity will be sought for a family of mappings, $\Gcal_\Omega \defeq \{\Gcal(x_0)\mid x_0\in\Omega \},$ defined over a robustly positive invariant subset of the state space, $\Omega \subset  \mathbb{R}^n$, for inputs in a non-empty set $\Ucal_\Omega\subseteq\Ell_{2}^m$ defined by
\begin{align*}
    \Ucal_\Omega \defeq& \bigcap_{x_0\in\Omega} \Ucal_\Omega(x_0) \\
    \Ucal_\Omega(x_0) \defeq& \left\{ \bu \in \Ell_{2e}^m \mid  x(t) \in \Omega\ \forall t\geq 0 \right\}.
\end{align*}



The following sections detail the necessary definitions and tools for local dissipativity analysis.

%

\subsection{Local Dissipativity}

Hill and Moylan informally defined a dissipative system as a ``system that absorbs more energy from the external world than it supplies" \cite{hill1976stability}. Essentially, any change in the system's internal energy (represented by the positive semi-definite ``storage function") is bounded above by a function of the input signal to the system and the resulting output, as shown below.

\begin{defn} \label{def:diss}  \cite[Definition 2]{willems1972dissipative}: A dynamical system \eqref{eq:dynSys} is dissipative with respect to a locally integrable supply rate $w(t) = w(u,y)$ if there exists a non-negative function, $V:\mathbb{R}^n\rightarrow\mathbb{R}_+$, called the storage function, such that for all $(t_0,t_1)\in\mathbb{R}_+$, $x \in \mathbb{R}^n$, and $\bu\in\Ell_2^m$
\begin{flalign}\label{eq:dissInequal}
    & V(x_0) + \int_{t_0}^{t_1}w(t)dt \geq V(x_1),
\end{flalign}
where $x_1$ and $y(t)$ are the result of applying \eqref{eq:dynSys} from time $t_0$ to $t_1$ with input $\bu(t).$
\end{defn}

While dissipativity was originally defined globally, there are many nonlinear systems for which dissipativity is a local property about an attractor \cite{hill1980connections,vidyasagar1982new}. Different definitions of local dissipativity have been used throughout the literature with various input signals, attractors, and system definitions \cite{hill1980connections,vidyasagar1982new,pota1993stability}. Below, \defref \ref{def:localDiss} gives a local QSR dissipativity that brings together several concepts in the literature \cite{hill1980connections,pota1993stability}.

\begin{defn}\label{def:localDiss}
    The mapping \eqref{eq:dynSys} is locally QSR dissipative on $\Omega \in\mathfrak{R}^n$, 
    if the following conditions are satisfied:
    \begin{enumerate} 
        \item The set $\Ucal_{\Omega}(x_0)$ 
        contains a neighborhood of the origin.
        \item Given the supply rate function 
        \begin{flalign}\label{eq:supplyRate}
            w(u,y) = y^\top\Q y + 2y^\top \bS u + u^\top \R u,
        \end{flalign}
        the inequality
        \begin{equation}\label{eq:dissInequal_local}
            \int_{0}^Tw(u,y) \geq \beta
        \end{equation}
        is satisfied for all $\bu\in\Ucal_{\Omega}(x_0)$ and $T\geq 0$, where $\beta\in\mathbb{R}$, $\Q = \Q^\top \vecdim{p\times p}$, $\bS \vecdim{p\times m}$, and $\R =\R^\top \vecdim{m\times m}.$
    \end{enumerate}
    The family of mappings $\Gcal_\Omega$ is locally QSR dissipative on $\Omega$ with inputs in $\Ucal_\Omega$ if all of its elements are dissipative with a shared supply rate and $\Ucal_\Omega$ contains a neighborhood of the origin.
\end{defn}

Sufficient conditions for global QSR dissipativity of smooth, nonlinear, control affine systems were formatted in \cite[Theorem 1]{hill1976stability}, \cite[Theorem 2]{van19922}. 
Local dissipativity can be established instead by applying the proof strategy from \cite[Theorem 2]{van19922} while restricting the initial state to a set, $x_0\in\Omega$, that is robustly positive invariant to inputs in $u\in\Ucal_\Omega\neq \emptyset$ and using a more general storage function. 
The resulting conditions 
are summarized next, providing a \ac{mi} that mirrors the \ac{kyp} Lemma.

\begin{thm}  \cite[Theorem 1]{hill1976stability}, \cite[Theorem 2]{van19922} \label{thm:dissConditions}
    Suppose that for some $\Omega \subset \mathbb{R}^n$ and dynamics defined by \eqref{eq:dynSys}, $\Ucal_{\Omega}$ contains a neighborhood of the origin. 
    If there exists a non-negative, smooth function, $V: \Omega \rightarrow \mathbb{R}_+,$ that satisfies 
    \begin{equation}\label{eq:qsrLMI}
        \bmat{\nabla V^\top f(x) & \frac{1}{2}\nabla V^\top \bar{G}(x)- h(x)^\top\bS & h(x)^\top \\ * & \hat{\R} & \bar{J}(x)^\top \\ * & * & \Q^{-1}} \preceq 0,
    \end{equation}
    where $\hat{\R} = -\R -\bS^\top \bar{J}(x) - \bar{J}(x)^\top\bS$, then $\Gcal_\Omega$ is locally dissipative on $\Omega$ with respect to \eqref{eq:supplyRate}.
\end{thm}
\begin{pf}
    First note that for each $x_0 \in \Omega$, $\Ucal_\Omega(x_0)$ contains a neighborhood of the origin because  $\Ucal_\Omega\subseteq\Ucal_\Omega(x_0)$. For $\bu \in \Ucal_\Omega(x_0),$ we can follow the steps proving \cite[Theorem 1]{hill1976stability} with $V(x) = \phi(x)$ up to \cite[Equation 6]{hill1976stability}, which verifies \eqref{eq:dissInequal_local}, and consequently local QSR dissipativity.  
\end{pf}

\subsection{Continuous Piecewise Affine Functions}

The storage function $V$ is non-unique and has no set functional form \cite{willems1972dissipative}. Further, the QSR  dissipativity condition is a function of $x \in \Omega$. \ac{cpa} functions will help us in the nontrivial task of searching for a storage function which satisfies \eqref{eq:qsrLMI} across $\Omega$. Inspired by the \ac{cpa} Lyapunov function search techniques of \cite{giesl2012construction,giesl2014revised}, we will use this flexible functional form for candidate storage functions.

\begin{defn} \cite{giesl2014revised} \textit{Affine independence}:
    A collection of $m$ vectors $\{x_0, x_1, \hdots , x_m\} \subset \mathbb{R}^n$ is affinely independent if $x_1-x_0, \hdots, x_m - x_0$ are linearly independent.  
\end{defn}
\begin{defn}\cite{giesl2014revised} \textit{$n$ - simplex}:
    A simplex, $\sigma$, is defined the convex hull of $n+1$ affinely independent vectors, $co\{x_j\}_{j=0}^n$, where each vector, $x_{j} \in \mathbb{R}^n$, is a vertex.
\end{defn}
\begin{defn}\cite{giesl2014revised} \textit{Triangulation}:
    Let $\Tcal {=} \{\sigma_i\}_{i=1}^{m_{\Tcal}} \in \mathfrak{R}^n$ represent a finite collection of $m_{\Tcal}$ simplexes, where the intersection of any two simplexes is a face or an empty set.
\end{defn}

Let $\Tcal = \{\sigma_i\}_{i=1}^{m_{\Tcal}}.$ Further, let $\{x_{i,j}\}_{j=0}^n$ be a simplex's ($\sigma_i$) vertices. The choice of $x_{i,0}$ in $\sigma_i$ is arbitrary unless $0 \in \sigma_i,$ in which case $x_{i,0} = 0$ \cite{giesl2014revised}. The vertices of the triangulation $\Tcal$ in a set $\Omega$ are denoted as $\mathbb{E}_{\Omega}.$
Let $\Tcal_{0}$ denote the simplexes in $\Tcal$ containing $0$ and $\Tcal_{\Omega\setminus\{0\}}$ denotes simplices that do not contain $0$.

An important characteristic of a \ac{cpa} function is that it is defined uniquely over \textit{all} of $\Omega \in\mathfrak{R}^n$ by its values on the \textit{finite set} of vertices in $\Tcal,$ as detailed in \lemref \ref{lem:lemma_gradW}.

%
\begin{lem}\label{lem:lemma_gradW} \cite[Remark 9]{giesl2014revised}
    Consider the triangulation $\Tcal = \{\sigma_i\}_{i=1}^{m_{\Tcal}},$ where $\sigma_i = \co(\{x_{i,j}\}_{j=0}^n)$, and a set $\W = \{W_{x}\}_{x \in \mathbb{E_{\Tcal}}}\subset\mathbb{R},$ where $W(x) = W_{x}, \forall x \in \mathbb{E}_{\Tcal}.$ For simplex $\sigma_i$, let $\X_i \vecdim{n \times n}$ be a matrix that has $x_{i,j} - x_{i,0}$ as its $j$-th row and $\bar{W}_i\vecdim{n}$ be a vector that has $W_{x_{i,j}} - W_{x_{i,0}},$ as its $j$-th element. The function $W(x) = x^\top \X_i^{-1}\bar{W}_i,$ is the unique \ac{cpa} interpolation of $\W$ on $\Tcal$ for $x \in \sigma_i$. 
\end{lem}


Like in \cite{giesl2012construction}, Taylor's Theorem is used to bound the change between a function $\zeta\in\mathcal{C}^2$ when evaluated at an arbitrary point versus the vertices on a simplex.


\begin{thm}[Taylor's Theorem \cite{fitzpatrick2009advanced}]\label{thm:taylor}
    Consider \\$\zeta:\map{n}{n}$, where $\zeta\in \mathcal{C}^2.$ For any $x,y \vecdim{n},$
    \begin{flalign}\label{eq:Taylor}
    \begin{split}
        \zeta^{(k)}(x) =& \zeta^{(k)}(y) {+} \langle \nabla \zeta^{(k)}  (x), x-y\rangle {+} \frac{1}{2}\overline{\bH}_{\zeta^{(k)}}(x,y),
    \end{split}
    \end{flalign}
    where $\overline{\bH}_{\zeta^{(k)}}(x,y) =  \langle \bH_{\zeta^{(k)}}(z) (x-y), (x-y)\rangle$ and $z$ is some convex combination of $x$ and $y.$
\end{thm}

Let $\overline{\bH}_{\zeta}(x,y) = \bmat{\overline{\bH}_{\zeta^{(1)}}(x,y) & \hdots & \overline{\bH}_{\zeta^{(n)}}(x,y)}^\top$ be known as the Taylor remainder.

Although storage functions must satisfy more complicated constraints than Lyapunov functions, the error bounds used to establish \ac{cpa} Lyapunov functions in \cite{giesl2012construction,giesl2014revised} will be useful here.

\begin{lem}\label{lem:lemma_eq_bound}  \cite[Proposition 2.2 and Lemma 2.3]{giesl2012construction}
Consider $\Omega \in \mathfrak{R}^n$. Let $\zeta{:} \Omega{ \rightarrow} \mathbb{R}^n$ where $\zeta \in \mathcal{C}^2(\Omega)$. Then, for any $x \in \sigma = \co(\{x_{j}\}_{j=0}^n) {\in \Omega}$,
\begin{flalign}\label{eq:infNormError}
    \norm{\zeta(x) {-} \!\!\sum_{j=0}^n\lambda_j \zeta(x_{j})}_{\infty} \!\!\!\!\!\!{=} &\frac{1}{2}\!\sumLam \!\norm{\overline{\bH}_\zeta(x,x_0) {-} \overline{\bH}_\zeta(x_j,x_0)}_\infty \nonumber \\   \!\!\!\!\!\leq &\frac{1}{2}n\beta\sum_{j=0}^n\lambda_j\bar{c}_{j},
\end{flalign}
where $\{\lambda_j\}_{j=0}^n \in \mathbb{R}_0^1$ is the set of unique coefficients satisfying $x = \sum_{j=0}^n\lambda_jx_{j}$ with $\sum_{j=0}^n\lambda_j = 1$,
\begin{flalign}
 & \beta \geq \max_{p,q,r \in\mathbb{Z}_1^n} \max_{\xi \in \sigma} \absVal{\frac{\partial^2\zeta^{(p)}(x)}{\partial x ^{(q)}\partial x ^{(r)}}\Bigr|_{x = \xi}}, \label{eq:oldB} \text{ and }\\
 &\bar{c}_{j} {=} \norm{x_{j} {-} x_{0}}\!(\max_{k \in \mathbb{Z}_1^n}\norm{x_{k} {-} x_{0}}\! {+} \!\norm{x_{j} {-} x_{0}}). \label{eq:oldC}
\end{flalign}
\end{lem}

In \axref \ref{sec:fxnError},  \lemref \ref{lem:fxnBound} establishes a similar, but tighter, error bound than the one presented above. The proof also uses Taylor's Theorem to bound the change in a function across a simplex. But while \lemref \ref{lem:lemma_eq_bound} performs a Taylor expansion about a vertex point of the simplex, $x_0$, \lemref \ref{lem:fxnBound} (\axref \ref{sec:fxnError}) performs the expansion about an arbitrary point within the simplex. Importantly, the behavior of these error bounds differs at the origin. When $x_j = 0$, $c_j = 0$ in \lemref \ref{lem:lemma_eq_bound}, while $c_j \neq 0$ in \lemref \ref{lem:fxnBound}. Thus, we will use the tighter bound from \lemref \ref{lem:fxnBound} throughout, except when simplices contain the origin. Then, we will revert to using \lemref \ref{lem:lemma_eq_bound}.

%

\section{Error bounds for function synthesis}\label{sec:ErrorBounds}

\lemref \ref{lem:lemma_eq_bound} is an invaluable tool for ensuring that \ac{cpa} function adheres to a scalar inequality condition. In this section, we expand this tool to more complex \ac{mi} conditions, which often codify complex control analysis and synthesis requirements. 
We develop novel error bounds for an \ac{mi} on an n-simplex and later leverage them for dissipativity analysis. 

In \secref \ref{sec:lmiBound2}, an additional \ac{lmi} is developed to enforce a scalar inequality constraint over a closed ball about the origin. This is later used to combine \ac{cpa} and quadratic storage functions for dissipativity analysis. 

\subsection{LMI Error Bounds}\label{sec:lmiBound1}

A positive definite error bound matrix is developed for a \ac{mi} constraint applied to an $n-$simplex. Enforcing the \ac{mi} constraint plus its error bound on the vertex points of an $n$-simplex ($\sigma=\co\{x_j\}_{j=0}^n$) implies that the \ac{mi} holds for all points within that simplex ($x \in \sigma$). While \cite{giesl2012construction} established \lemref \ref{lem:lemma_eq_bound} as a general error bound for $\mathcal{C}^2$ vector-valued functions, it does not translate automatically to a \ac{mi} that contains $\mathcal{C}^2$ vector-valued functions because the structure of a \ac{mi} affects its definiteness. This factor is considered in the following theorem.

\begin{thm}\label{thm:lmi_bound}
    Consider
    \begin{flalign}\label{eq:genericMat}
        \M(x) = \bmat{\phi(x) & \0 \\ \0 & -\I} + \text{He}\left(\bmat{0 &\zeta(x)^\top \\ \0 & \0}\bmat{0 & \0 \\ \0 & \I}\right),
    \end{flalign}
     where $x \vecdim{n},$ $\phi:\map{n}{},$ $\zeta:\map{n}{m},$ $\phi,\zeta \in \mathcal{C}^2.$ Let $\sigma \defeq \co\{x_j\}_{j=0}^n$ be an n-simplex in $\mathbb{R}^n.$ If $x = \sum_{j=0}^n\lambda_jx_j \in \sigma,$ then
    \begin{flalign}\label{eq:lmi_boundSchur}
        \M(x) - &\sum_{j=0}^n \lambda_j \M(x_j) \preceq  \nonumber \\ &
        \sum_{j=0}^n \lambda_j \bmat{\frac{1}{2}\hat{\phi}_j & *  & * \\ \0 & \frac{1}{2}\Pi^{-1} & * \\ \hat{\zeta} & \0 & -2\Pi^{-1}} \defeqRight \E(x),
    \end{flalign}
    where $\hat{\phi}_j \geq \max_{j\in\mathbb{Z}_0^n} \absVal{\overline{\bH}_\phi(x_j,x)}$, $\hat{\zeta} \vecdim{m\times 1}$ where each element $\hat{\zeta}^{(k)} \geq \max_{j\in\mathbb{Z}_0^n}\absVal{\overline{\bH}_{\zeta^{(k)}}(x_j,x)}$ for $k \in \mathbb{Z}_1^m$, and 
    $\Pi\in \mathbb{R}^{m \times m}$ is a diagonal, positive definite matrix with elements $\pi_k,$ $k \in \mathbb{Z}_1^m.$
    
    Moreover, if $\M(x_j) + \E(x_j) \preceq 0$ holds for all vertex points of $\sigma,$ then $\M(x) \preceq 0$ holds for all $x \in \sigma.$
\end{thm}
\thmref \ref{thm:lmi_bound} bounds the difference between $\M(x)$ at any convex combination of vertex points, $\M(\sum_{i=0}^n\lambda_i\x_i)$, 
and the convex combination of it evaluated at each vertex point, $\sum_{{j}=0}^n\lambda_{j}\M(x_{j})$. 
The proof parallels \cite[Proposition 2.2]{giesl2012construction} by developing remainder terms using Taylor's Theorem, but exploits the structure of $\M(x)$ to establish an \ac{lmi} error bound. Note the negative identity matrix in (\ref{eq:genericMat}) is essential to enforce negative semi-definiteness.

\begin{pf}
    Let $\tilde{\E}(x) = \M(x) - \sumLam \M(x_j).$ As in Lemma \ref{lem:fxnBound}, apply a \thmref \ref{thm:taylor} to $\phi$ and $\zeta$ about $x$ for each vertex point to get 
    \begin{flalign*}
        &\frac{1}{2}\sum\limits_{{j}=0}^n\lambda_{j}\!\!
        \bmat{{-} \overline{\bH}_{\phi}(x_j,x) & \0\\
        \0 & \0}\!\! {+} \text{He}\!\!\left(\! \bmat{\0 & -\overline{\bH}_\zeta(x_j,x)^\top\\ \0 & \0}\!\!\bmat{\0 & \0 \\ \0& \I}\!\right)\!\!,
    \end{flalign*}
    which we define as $\tilde{\E}(x).$ If $\tilde{\E}(x)$ was diagonal, bounding  $\tilde{\E}(x)$ across a simplex would be as simple as bounding its individual elements. Since that is not the case, the structure of the matrix is leveraged to bound its off-diagonal terms.

    Young's relation  \cite{zhou1988robust} is used, with scaling matrix $\bar{\Pi} = \bmat{\I & \0 \\ \0 & \Pi}$, to bound the Hermition of $\tilde{\E}(x)$, resulting in
    \begin{flalign*}
        \tilde{\E}(x) & \preceq \frac{1}{2}\sum\limits_{{j}=0}^n\lambda_{j}
        \bmat{ e_j& \0\\
        \0 & \Pi^{-1}},
    \end{flalign*}
    where $e_j =  {-}\bar{\bH}_{\phi}(x_j,x)  + \bar{\bH}_{\zeta}(x_j,x)^\top \Pi \bar{\bH}_{\zeta}(x_j,x).$ Note that the structure of $\X$ ensures that  no cross terms are formed between individual elements of $\bar{\bH}_\zeta$ in $\e_j$. Now each element can be individually bounded.
    Here, $e_j \leq \hat{\phi}_j + \hat{\zeta}^\top \Pi \hat{\zeta}.$ Perform a Schur complement \cite{boyd1994linear} about the term $\hat{\zeta}^\top \Pi \hat{\zeta}$ to get \eqref{eq:lmi_boundSchur}.

    Now suppose that $\M(x) \preceq 0$ must be imposed on all $x \in \sigma.$ By assumption, $\M(x_{j}) + \E(x_{j}) \preceq 0$ holds for each vertex of $\sigma$  $(x_0, \hdots, x_n).$ The set of negative semi-definite \acp{lmi} is a convex cone \cite{boyd2004convex}. By enforcing $\M(x_{j}) + \E(x_{j}) \preceq 0$ on each vertex, the expression $\sum_{{j}=0}^n\lambda_{j}(\M(x_{j}) + \E(x_{j})) \preceq 0$ also holds. The \ac{lmi} $\M(x) + \E(x) \preceq 0$ implies $\M(x)\preceq 0,$ because $\E(x)\succeq 0$. Therefore, $\M(x) \preceq 0$ for all $x \in \sigma.$\qed
\end{pf}

The weighing matrix $\Pi$ in \eqref{eq:lmi_boundSchur} of \thmref \ref{thm:lmi_bound} allows for a user to have some control over the error bound of the off diagonal terms of $\M(x)$ and can be treated as a design variable in an optimization problem.

In \thmref \ref{thm:lmi_bound}, functions contained within the \ac{mi} are linearized about an arbitrary point within a simplex  ($x \in\sigma$). However, a similar process to \lemref \ref{lem:lemma_eq_bound} can used to linearize the functions about a fixed vertex, $x_0,$ resulting in the corollary stated in \axref \ref{sec:lmiBoundOrigin}. 

\subsection{Quadratic Error Bounds}\label{sec:lmiBound2}

\secref \ref{sec:lmiBound1} bounded matrix behavior across a simplex of a triangulation. 
Here, the focus shifts to bounding an inequality constraint within a closed two-norm ball about the origin. As in previous sections, Taylor's Theorem bounds the inequality across the region. However, by focusing on a two-norm ball about the origin, \thmref \ref{thm:originLMI} expands concepts used in \ac{cpa} function synthesis to different function types. 

\begin{thm}\label{thm:originLMI}
    Consider the inequality
    \begin{equation}\label{eq:genIneqal}
         \zeta(x)^\top\zeta(x)+ \frac{1}{2}\left(x^\top \theta(x) + \theta(x)^\top x\right) \leq 0,
    \end{equation}
    where $x \vecdim{n}$, $\theta:\mathbb{R}^n\rightarrow\mathbb{R}^n,$ $\zeta:\mathbb{R}^n\rightarrow\mathbb{R}^m,$ $\theta, \zeta \in \mathcal{C}^2,$ $\theta(0)  {=}0, $ and $\zeta(0) {=} 0.$ Let $\hat{\theta} {\geq} \max_{k \in \mathbb{Z}_1^n,\xi \in B_\epsilon(0)} \norm{\bH_{\theta^{(k)}}(\xi) }_2$ and $\hat{\zeta} {\geq} \max_{k \in \mathbb{Z}_1^m,\xi \in B_\epsilon(0)} \norm{\bH_{\zeta^{(k)}}(\xi) }_2$. Let $\mathbf{J}_{\theta}$ and $\mathbf{J}_{\zeta}$ be the Jacobians of $\theta$ and $\zeta$ evaluated at $x = 0$.
    Further, let $\Pi \vecdim{m\times m}$ be a diagonal, positive definite matrix, where $\underline{\pi}$ is the minimal non-zero value of $\Pi^{-1}$. If 
    \begin{flalign}\label{eq:genEpLMI}
        \M_{\epsilon} {=} \bmat{\frac{1}{2}(\mathbf{J}_{\theta}\!^\top\!\!{+} \mathbf{J}_{\theta}{+}\hat{\theta}\epsilon\I)& \mathbf{J}_\zeta ^{\top} & \hat{\zeta}\epsilon\1_{1\times m} \\ * & -\I +\frac{1}{2}\Pi^{-1} & \0 \\ * & * & -2\underline{\pi}\I} \preceq 0
    \end{flalign}
    holds, then \eqref{eq:genIneqal} holds for all $x \in B_{\epsilon}(0)$.
\end{thm}

\begin{pf}
    For vector-valued functions, Taylor's Theorem can be applied to each dimension. By applying Taylor's Theorem about the origin,
    \begin{flalign*}
    \zeta(x)=& \mathbf{J}_{\zeta}x {+} \frac{1}{2}\bmat{x^\top \bH_{\zeta^{(1)}}(z_{\zeta^{(1)}})\\ \vdots \\ x^\top\bH_{\zeta^{(m)}}(z_{\zeta^{(m)}})}x {\defeqRight} (\A_1 {+} \frac{1}{2}\A_2)x \\
    \text{and likewise} &\\
    \theta(x)=&\mathbf{J}_{\theta}x{+} \frac{1}{2}\bmat{x^\top\bH_{\theta^{(1)}}(\z_{\theta^{(1)}})\\ \vdots \\ x^\top\bH_{\theta^{(n)}}(z_{\theta^{(n)}})}x{ \defeqRight} (\B_1 {+} \frac{1}{2}\B_2)x,
    \end{flalign*}
    where $\A {=} \A_1 {+} \frac{1}{2}\A_2$, $\B {=} \B_1 {+} \frac{1}{2}\B_2$. Then, \eqref{eq:genIneqal} is equivalently expressed as $x^\top\A^\top\A x + \frac{1}{2}(x^\top\B x{+}x^\top\B^\top x) \leq 0$. 
    
    By factoring out $x^\top$ and $x$ and performing a Schur complement \cite{boyd1994linear} on $\A^\top\A$, \eqref{eq:genIneqal} becomes the \ac{lmi}
    \begin{equation*}
        \bmat{\frac{1}{2}(\B{+}\B^\top) & \A_1^\top \\ \A_1&-\I } + \text{He}\left(\bmat{\0 &\0 \\ \0 & \I}^\top \bmat{\0 & \0 \\ \frac{1}{2}\A_2 & \0}\right)\preceq 0.
    \end{equation*}
    Both $\A_2$ and $\B_2$ contain $x$ -- meaning that an infinite number of constraints would be needed to enforce the \ac{lmi} for all $x \in B_{\epsilon}(0).$ Therefore, the constraint can instead be enforced by a singular \ac{lmi} that bounds it above for all $x \in B_{\epsilon}(0).$

    First, Young's relation, with scaling matrix $\bar{\Pi} = \bmat{\pi_0 & \0 \\ \0 & \Pi}$ ($\pi_0 > 0)$, bounds the Hermition above to produce the \ac{lmi}
     \begin{equation*}
        \bmat{\frac{1}{2}(\B{+}\B^\top + \A_2^\top\Pi\A_2) & \A_1^\top \\ \A_1&-\I +\frac{1}{2}\Pi^{-1} } \preceq 0.
    \end{equation*}
    Then, the definition of matrix two norms and the Cauchy Schwarz inequality are used to produce the bounds $\A_2^\top\Pi\A_2 \preceq \norm{\A_2}_2^2\norm{\Pi}_2\I \preceq \hat{\zeta}^2\epsilon^2\norm{\Pi}_2\I$ and $\B_2 \preceq \hat{\theta}\epsilon \I$. Finally, perform a Schur complement \cite{boyd1994linear} on $\frac{1}{2}\hat{\zeta}^2\epsilon^2\norm{\Pi}_2\I$, which results in the term $-\frac{2}{\norm{\Pi}}$ on the diagonal of the matrix. Let $\pi_M$ be the maximum term of $\Pi$. For a diagonal, square matrix $\norm{\Pi}_2 \leq \pi_M$ and therefore, $-\frac{2}{\norm{\Pi}_2} \leq -\frac{2}{\pi_M}.$ Recall that $\frac{1}{\pi_M}$ is the minimum, nonzero term of $\Pi^{-1}$ to get \eqref{eq:genEpLMI}. Thus, \eqref{eq:genEpLMI} implies \eqref{eq:genIneqal}. \qed
\end{pf}

As in the previous section, it may be desirable in some instances to treat $\Pi$ in \eqref{eq:genEpLMI} as a design variable.

\section{Local dissipativity analysis}\label{sec:dissAnalysis}

The error bounds developed in  \secref \ref{sec:ErrorBounds} are now leveraged to formulate conditions for which a \ac{cpa} function or combined \ac{cpa} and quadratic function is the storage function of a dissipative system. \secref \ref{sec:opt} then uses these to develop convex optimization problems that simultaneously synthesize a system's storage function and analyze common QSR characterizations -- such as gain, passivity, and conic bounds.

\subsection{Storage function conditions}\label{sec:qsrLMI}

\thmrefs \ref{thm:dissNoCA} and \ref{thm:dissCA} give conditions for a \ac{cpa} function or a combined \ac{cpa} and quadratic function, respectively, adhere to the QSR dissipativity condition \eqref{eq:qsrLMI} across a region of the state space. \thmref \ref{thm:dissNoCA} considers the case where the system dynamics make all input affine terms disappear at the origin, while \thmref \ref{thm:dissCA} admits non-zero control affine terms at the origin. This distinction is important, as all error bounds in the optimization problem must be zero at the origin to prevent infeasibility. While this always occurs when the input term disappears, a modified quadratic and \ac{cpa} storage function (which is quadratic at the origin) is needed for the nonzero case.  

\begin{thm}\label{thm:dissNoCA}
    Consider \eqref{eq:dynSys} with $\B = 0$ and $\D = 0$. Suppose that $f,G,h,J \in \mathcal{C}^2$ on the set $\Xcal \in \mathfrak{R}^n$  triangulated by $\Tcal.$  Let $\Q = \Q^\top \vecdim{p\times p}$, $\bS \vecdim{p\times m}$, and $\R=\R^\top \vecdim{m\times m},$ where $\Q \prec 0$ and $\R \succ 0$. Let $\Pi_1 \vecdim{m\times m}$ and $\Pi_2,\Pi_3 \vecdim{p \times p}$ be diagonal, positive definite matrices, where $\underline{\pi}_{z}$ is the minimum diagonal value of $\Pi_z^{-1}$. Define $\mathbf{L} = \{l_i\}_{i=1}^{m_{\Tcal}}$. Define the \ac{cpa} function, $V = \{V_x\}_{x\in\mathbb{E}_{\Tcal}}$. Consider
    \begin{subequations}\label{eq:dissipativityOptA}
        \begin{flalign}
            V_x &\geq 0 \quad \forall x \in \mathbb{E}_\Tcal \\
            \norm{\nabla V_i}_1 &\leq l_i \quad \forall i \in \mathbb{Z}_{1}^{m_{\Tcal}}, \label{eq:boundVNorm}\\
            \norm{\bS}_2 &\leq \hat{s}, \\
            \M(x_{i,j}) + \overline{\E}(x_{i,j}) &\preceq 0, \quad \forall i \in \mathbb{Z}_{1}^{m_{\Tcal}}, \forall j \in \mathbb{Z}_0^n, \label{eq:constraint_QSRLMI_noCA} \\ & \quad \quad \quad x_{i,j} \neq 0 \nonumber
        \end{flalign}    
    \end{subequations}
    where $\M$ is the QSR dissipativity condition \eqref{eq:qsrLMI} and
     \begin{flalign}\label{eq:lmiNoControlAffine}
     \begin{split}
        \overline{\E}(x_{i,j}) &= \\ & \bmat{\hat{f}_{i,j} & *&* &*&*&*&* \\
          \0 & \hat{\R} & * & *& * &*&*\\
         \0 & \0 & \hat{\Q} & * & *  &*&*\\
          \hat{s} \hat{h}_{i,j} & \0 &\0  & -2\underline{\pi}_{1}\I &*&*&*\\
          l_i\hat{G}_{i,j} & \0 & \0 & \0 & -4\underline{\pi}_1\I&*&*\\
         \hat{h}_{i,j}  &\0 &\0 & \0 & \0 & -2\Pi_2^{-1}&*\\
         \0 & \hat{J}_{i,j} & \0 & \0 &\0 &\0 &\!\!\! -2\underline{\pi}_3\I
         },
     \end{split}
     \end{flalign}
    where $\hat{\R} = \frac{1}{2}(\Pi_1^{-1} {+} \hat{s}n\sqrt{pm}\theta_i c_{i,j} \I)$,  $\hat{\Q} = \frac{1}{2}(\Pi_2^{-1} {+} \Pi_3^{-1})$, $\hat{f}_{i,j} = \frac{n}{2}l\beta_i c_{i,j}$, $\hat{G}_{i,j} = n\bmat{\mu_{1}c_{i,j} & \hdots & \mu_{m}c_{i,j}}^\top$, $\hat{h}_{i,j} = n\bmat{\rho_{1}c_{i,j} & \hdots & \rho_{p}c_{i,j}}^\top$, and $\hat{J}_{i,j} = n\sqrt{pm}\theta_ic_{i,j}\I.$
    Further, $c_{i,j}$ equals $\overline{c}_j$ in \eqref{eq:oldC} when $0 \in \sigma$ and follows \eqref{eq:c} when $0 \notin \sigma,$  
    \begin{flalign}\label{eq:beta}
        \beta_i \geq \max_{p,q,r \in \mathbb{Z}_1^n, \xi\in\sigma_i} \absVal{\frac{\partial f^{(p)}}{\partial x ^{(q)} \partial x^{(r)}}\Big|_{x =\xi}},
    \end{flalign}
    \begin{flalign}\label{eq:rho}
        \rho_{i,a} \geq \max_{q,r \in \mathbb{Z}_1^n, \xi\in\sigma_i} \absVal{\frac{\partial h^{(a)}}{\partial x ^{(q)} \partial x^{(r)}}\Big|_{x =\xi}},
    \end{flalign}
    \begin{flalign}\label{eq:mu}
        \mu_{i,k} \geq \max_{p,q,r \in \mathbb{Z}_1^n, \xi\in\sigma_i} \absVal{\frac{\partial G_k^{(p)}}{\partial x ^{(q)} \partial x^{(r)}}\Big|_{x =\xi}},
    \end{flalign}
    and
    \begin{flalign}\label{eq:theta}
        \theta_{i} \geq \max_{k \in \mathbb{Z}_1^m,a \in \mathbb{Z}_1^p, q,r \in \mathbb{Z}_1^n, \xi\in\sigma_i} \absVal{\frac{\partial J_k^{(a)}}{\partial x ^{(q)} \partial x^{(r)}}\Big|_{x =\xi}}.
    \end{flalign}

   If $V$ satisfies \eqref{eq:dissipativityOptA}, then $V$ satisfies the dissipation condition \eqref{eq:qsrLMI} for all $x \in \Xcal^\circ$ for $\Q,$ $\bS$, and $\R.$
\end{thm}

\begin{pf}
    \thmref \ref{thm:lmi_bound} is applied to the QSR dissipativity condition \eqref{eq:qsrLMI} resulting in Constraint \eqref{eq:constraint_QSRLMI_noCA}, which ensures that  \eqref{eq:qsrLMI} holds over each simplex for $V$. To expand, the \eqref{eq:qsrLMI} is equivalently expressed as
    \begin{flalign*}
         \M(x) & =\bmat{\nabla V^\top f(x) &\0 & \0 \\ * & \hat{\R} & \0 \\ * & * & \Q^{-1}} \\ &+ \text{He}\left(\bmat{0 & \frac{1}{2}\nabla V^\top G(x){-}h(x)^\top\bS & \0 \\ \0 &\0 &\0 \\ \0& \0 & \0}\bmat{0 &\0 & \0 \\ \0 & \I & \0 \\ \0 & \0 & \0}\right) \\ 
        &+ \text{He}\left(\bmat{0 & \0 & h(x)^\top \\ \0 &\0 &\0 \\ \0& \0 & \0}\bmat{0 &\0 & \0 \\ \0 & \0 & \0 \\ \0 & \0 & \I}\right) \\
        &+ \text{He}\left(\bmat{0 &\0 & \0 \\ \0 &\0 &J(x)^\top \\ \0& \0 & \0}\bmat{0 &\0 & \0 \\ \0 & \0 & \0 \\ \0 & \0 & \I}\right).
    \end{flalign*}
   Consider the simplex $\sigma_i$, where $0 \notin \sigma_i$. Therefore, for any $x \in \sigma_i,$ the bound on the \ac{lmi} evaluated within the simplex versus on the vertices, i.e. $\M(x) - \sumLam \M(x_{i,j})$, is
   \begin{flalign*}
       \E(x) = \bmat{\Xi_1 & \0 & \0 \\ \0 & \Xi_2 + \frac{1}{2} \Pi_1^{-1} & \0 \\ \0 & \0 & \frac{1}{2}(\Pi_2^{-1} + \Pi_3^{-1})},
   \end{flalign*}
   where 
   \begin{flalign*}
       \Xi_1  \geq \frac{1}{2}\sumLam&\Bigl(\!\nabla V^\top \overline{\bH}_{f} (x_{i,j},x) \\ &  + \bigl(\frac{1}{2}\nabla V^\top \overline{\bH}_G(x_{i,j},x) {-} \overline{\bH}_h(x_{i,j},x)^\top \bS\bigr)\Pi_1\\ &\bigl(\frac{1}{2}\overline{\bH}_G(x_{i,j},x)^\top \nabla V-\bS \overline{\bH}_h(x_{i,j},x)\bigr) \\ & + \overline{\bH}_h(x_{i,j},x)^\top \Pi_2 \overline{\bH}_h(x_{i,j},x)\Bigr)
   \end{flalign*} 
   and 
   \begin{flalign*}
       \Xi_2 \succeq \sumLam &\Bigl(\bS^\top \overline{\bH}_J(x_{i,j},x)\\ & + \frac{1}{2} \overline{\bH}_J(x_{i,j},x)^\top \Pi_3  \overline{\bH}_J(x_{i,j},x)\Bigr).
   \end{flalign*}

   To bound Taylor remainder terms, $\overline{\bH}$, used in $\Xi_1$ and $\Xi_2$, \lemref \ref{lem:fxnBound} is used. Care must first be taken to get these functions into the form where the lemmas can be applied.

   First, consider $\Xi_1.$ Bounding $\frac{1}{2}\sumLam\nabla V^\top \overline{\bH}_{f} (x_{i,j},x)$ follows the procedure of \cite{giesl2014revised}. By the linearity of $\nabla V$ and H\"older's inequality, the expression is bounded above by \\ $\frac{1}{2}\norm{\nabla V}_1\sumLam\norm{\overline{\bH}_{f} (x_{i,j},x)}_\infty$. From here, \lemref \ref{lem:fxnBound} is be applied. To bound $\frac{1}{2}\sumLam (\frac{1}{2}\nabla V^\top \overline{\bH}_G(x_{i,j},\!x)$\\$- \overline{\bH}_h(x_{i,j},x)^\top \bS)\Pi_1(\frac{1}{2}\overline{\bH}_G(x_{i,j},x)^\top \nabla V-\bS \overline{\bH}_h(x_{i,j},x)),$ apply the Cauchy-Schwarz inequality, the triangle inequality, and the properties of matrix operator norms to 
   get $\frac{1}{2}\norm{\Pi}_1\sumLam(\norm{\overline{\bH}_{h}(x_{i,j},x)}_2^2\norm{\bS}_2^2$\\$ + \frac{1}{2}\norm{\nabla V^\top \overline{\bH}_G(x_{i,j},x)}_2^2).$ Note also that\\  $\norm{\nabla V^\top \overline{\bH}_G(x_{i,j},x)}_2^2 \leq \norm{\nabla V}_1^2\sum_{k=1}^m\norm{\overline{\bH}_{G_k}(x_{i,j},x)}_\infty^2$. From here, \lemref \ref{lem:lemma_eq_bound} is applied to terms involving $h$ and $G.$ Finally, $ \overline{\bH}_h(x_{i,j},x)^\top \Pi_2 \overline{\bH}_h(x_{i,j},x)$\\$ = \sum_{k=1}^p\overline{\bH}_{h^{(k)}}(x_{i,j},x)^2\pi_{2,k}$ can be bounded using the same procedure as in \lemref \ref{lem:fxnBound}. Therefore,
   \begin{flalign*}
       \Xi_1 =& \sumLam \Bigl( \hat{f}_{i,j} + \frac{1}{2}\hat{h}_{i,j}^\top \Pi_2 \hat{h}_{i,j} \\& + \frac{1}{2} \norm{\Pi_1}_2 \hat{s}^2 \hat{h}_{i,j}^\top \hat{h}_{i,j} + \frac{1}{4}\norm{\Pi_1}_2l_i^2 \hat{G}_{i,j}^\top\hat{G}_{i,j} \Bigl).
   \end{flalign*}

    To determine $\Xi_2,$ use the properties of matrix norms to produce the bound 
   \begin{flalign*}
       &\sumLam (\bS^\top \overline{\bH}_{J_{i,j}} + \frac{1}{2}\overline{\bH}_{J}(x_{i,j},x)^\top \Pi_3 \overline{\bH}_{J}(x_{i,j},x)) \preceq \\ & \sumLam (\norm{\bS}_2 \norm{\overline{\bH}_{J}(x_{i,j},x)}_2 + \frac{1}{2}\norm{\overline{\bH}_{J}(x_{i,j},x)}_2^2\norm{\Pi_3}_2)\I.
   \end{flalign*}
   Thus,
   \begin{flalign*}
       \Xi_2 = & \sumLam \Bigl(\hat{s}n\sqrt{pm}\theta_ic_{i,j} +\frac{1}{2}\hat{J}_{i,j}^2\norm{\Pi_3}_2\Bigr)\I.
   \end{flalign*}

    Schur complements \cite{boyd1994linear} are performed about \\ $\frac{1}{2} \norm{\Pi_1}_2 \hat{s}^2 \hat{h}_{i,j}^\top \hat{h}_{i,j},$ $\frac{1}{4}\norm{\Pi_1}_2l_i^2 \hat{G}_{i,j}^\top\hat{G}_{i,j}$,  $ \frac{1}{2}\hat{h}_{i,j}^\top \Pi_2 \hat{h}_{i,j}$,  and $\frac{1}{2}\hat{J}_{i,j}^2\norm{\Pi_3}_2\I$ to get \eqref{eq:lmiNoControlAffine}, where $\frac{1}{\norm{\Pi}}$ is equivalent to the minimal diagonal value of $\Pi^{-1}.$ 

    When $x \in \sigma_i$, where $0 \in \sigma_i$, the same procedure can be applied using \corref \ref{cor:lmi_bound_gh} rather than \thmref \ref{thm:lmi_bound}. Then, bounds on Taylor remainder terms, $\overline{\bH}$ are found using \lemref \ref{lem:lemma_eq_bound} rather than \lemref \ref{lem:fxnBound}. The only difference between Constraint \eqref{eq:constraint_QSRLMI_noCA} when found using this method is that $c_{i,j}$ will be defined using \eqref{eq:oldC} rather than \eqref{eq:c}.
    
    The values \eqref{eq:beta}, \eqref{eq:rho}, \eqref{eq:mu}, and \eqref{eq:theta} exist for all simplexes, because $f, G, h,J \in \mathcal{C}^2$ and $\sigma \subseteq \Xcal$ is bounded. Furthermore, $c_{i,j}$ is finite on $\sigma,$ because $\Xcal \in \mathfrak{R}^n.$ Defining $c_{i,j}$ as in \eqref{eq:oldC} when $0 \in \sigma$ ensures that the error bound is $0$ at the origin.
    
    By assumption, \eqref{eq:dissipativityOptA} holds on each $n$-simplex of the $\Tcal$. By \thmref \ref{thm:lmi_bound}, \eqref{eq:qsrLMI} holds for all $x \in\Xcal^\circ$ for some $\Q,$ $\bS$, and $\R.$\qed 

\end{pf}



\begin{thm}\label{thm:dissCA}
    Consider \eqref{eq:dynSys}. Suppose that $f,G,h,J \in \mathcal{C}^2$ on the set $\Xcal \in \mathfrak{R}^n.$  Let $\Q = \Q^\top \vecdim{p\times p}$, $\bS \vecdim{p\times m}$, and $\R=\R^\top \vecdim{m\times m},$ where $\Q \prec 0$ and $\R \succ 0$. Let $\Pi_1, \Pi_4,\Pi_5, \Pi_6 \vecdim{m\times m}$ and $\Pi_2,\Pi_3,\Pi_7,\Pi_8, \Pi_9 \vecdim{p \times p}$ be diagonal, positive definite matrices, where $\underline{\pi}_{z}$ is the minimum diagonal value of a matrix $\Pi_z^{-1}$. Consider the sets $0\in \Psi$ and $B_\epsilon(0)$, where $\Psi \subset B_\epsilon(0) \subset \Xcal.$ Define a triangulation $\Tcal{=}\{\sigma_i\}_{i=1}^{m_\Tcal}$ on $\Xcal\setminus \Psi^\circ$. Let there exist a \ac{cpa} function $V_C {=} \{\bar{V}_x\}_{x \in \mathbb{E}_{\Tcal}}$ satisfying \eqref{eq:dissipativityOptA} and a quadratic function $V_P = x^\top \bP x$ satisfying
    \begin{subequations}\label{opt1b:dissipativty}
        \begin{flalign}
            \bar{V}_x & \geq x^\top \bP x \quad \forall x \in \mathbb{E}_{\delta \Psi} \label{eq:Vc_higher}\\
            \bar{V}_x & \leq x^\top \bP x \quad \forall x \in \mathbb{E}_{B_\epsilon(0)\setminus \Psi} \label{eq:Vc_lower}\\
            \bP & \succ 0, \\
            \norm{\bP}_2 & \leq l_p, \\
            \M_{\epsilon} & \preceq 0, \label{eq:normBallConstr}
        \end{flalign}
    \end{subequations}
     where 
     \begin{flalign}\label{eq:lmiControlAffine}
        & \M_{\epsilon}= \nonumber\\
        & \!\!\!\bmat{\Xi_1 & \!\!\!* &* & * & * & * & *& * & *\\ 
         \!\!\! \B^\top\bP {-} \bS^\top \J_h & \!\!\!\Xi_2 & * &  * & * & * & *& * & *\\
        \J_h & \!\!\!\!\!\D & \Xi_3 & * & * & * & *& * & *\\
        \bar{\J}_G &\!\!\!\0 & \0 & \!\!\!-\underline{\pi}_4\I & *& * & *& * & *\\
        \hat{s}\epsilon n p^{\frac{1}{2}}\rho_\epsilon \I &\!\!\!\0 &\0 & \0 & \!\!\!-2\underline{\pi}_5\I & * & *& * & *\\
        l_pm^{\frac{1}{2}}n^{\frac{3}{2}}\epsilon^2\mu_\epsilon \I &\!\!\! \0 & \0 &\0\!\!\! &\0  &\! \!\!\!-2\underline{\pi}_6\I & * & * & *\\
        \epsilon n p^{\frac{1}{2}}\rho_\epsilon \I & \!\!\!\0 & \0 &\0 &\0 &\0 & \!\!\!\!-2\underline{\pi}_7\I & * & *\\
        \0&\!\!\! \bar{\J}_J & \0 &\0 &\0 &\0 & \0 & \!\!\!\!-\underline{\pi}_8\I & *\\
         \0& \!\!\! \bar{\theta} & \0 &\0 &\0 &\0 & \0 & \0& \!\!\! \!-2\underline{\pi}_{9}\I}\!\!.
     \end{flalign}
     Here, $\J_q$ represents the Jacobian of function $q$ evaluated at $0$, $\Xi_1 =  \bP\J_f + \J_f^\top\bP + l_pn^{\frac{3}{2}}\beta_\epsilon \epsilon\I_n$, $\Xi_2 = -\R -\bS^\top\D -\D^\top\bS + \Pi_4^{-1} + \frac{1}{2} (\Pi_5^{-1} + \Pi_6^{-1})+ (2\hat{s}\epsilon \sqrt{\sum_{k=1}^m\norm{\J_{j_k}(0)}_2^2} + \hat{s}(mp)^{\frac{1}{2}}n\epsilon^2\theta_\epsilon)\I$, $\Xi_3 = \Q^{-1} + \frac{1}{2}(\Pi_7^{-1} + \Pi_{9}^{-1}) + \Pi_8^{-1}$, $\bar{\J}_G = l_p\epsilon \sqrt{\sum_{k=1}^m\norm{\J_{g_k}(0)}_2^2} \I_{n}$, $\bar{\J}_J =\epsilon \sqrt{\sum_{k=1}^m\norm{\J_{j_k}(0)}_2^2} \I $ $\bar{\theta} = m^\frac{1}{2}p^\frac{1}{2}n\theta_\epsilon\epsilon^2 \I$, 
    \begin{equation}\label{eq:betaEp}
        \beta_{\epsilon} \geq \max_{p,q,r \in\mathbb{Z}_1^n,\xi \in B_{\epsilon}(0)}\absVal{\frac{\partial^2f^{(p)}(x)}{\partial x ^{(q)}\partial x ^{(r)}}\Bigr|_{x = \xi}},
    \end{equation}
    \begin{equation}\label{eq:rhoEp}
        \rho_{\epsilon} \geq \max_{p \in\mathbb{Z}_1^p, q,r \in\mathbb{Z}_1^n, \xi \in B_{\epsilon}(0)}\absVal{\frac{\partial^2h^{(p)}(x)}{\partial x ^{(q)}\partial x ^{(r)}}\Bigr|_{x = \xi}},
    \end{equation}
    \begin{equation}\label{eq:muEp}
        \mu_{\epsilon} \geq \max_{k\in \mathbb{Z}_1^{m}, p,q,r \in\mathbb{Z}_1^n, \xi \in B_{\epsilon}(0)}\absVal{\frac{\partial^2G^{(p)}_{k}(x)}{\partial x ^{(q)}\partial x ^{(r)}}\Bigr|_{x = \xi}},
    \end{equation}
    and 
    \begin{equation}\label{eq:thetaEp}
        \theta_{\epsilon} \geq \max_{k\in \mathbb{Z}_1^{m}, p \in \mathbb{Z}_1^p q,r \in\mathbb{Z}_1^n, \xi \in B_{\epsilon}(0)}\absVal{\frac{\partial^2J^{(p)}_{k}(x)}{\partial x ^{(q)}\partial x ^{(r)}}\Bigr|_{x = \xi}}.
    \end{equation}
    Further, $\beta_i, \rho_{i,a}$ and $\mu_{i,k}$ are defined by \eqref{eq:beta}, \eqref{eq:rho}, and \eqref{eq:mu}, and $c_{i,j}$ is defined by \eqref{eq:c}. Then
    \begin{flalign}\label{eq:V_cA}
        V(x) = \begin{cases}
            V_P(x) &\quad x \in \Psi\\
            \min \bmat{V_P(x) & V_C(x)} &\quad x \in B_\epsilon(0)\setminus \Psi\\
            V_C(x) &\quad x \in \Xcal \setminus B_\epsilon(0),\\
        \end{cases}
    \end{flalign}
    satisfies the dissipation condition \eqref{eq:dissInequal} for all $x \in \Xcal^\circ$ for $\Q$, $\bS$, and $\R.$

\end{thm}

\begin{pf}
    %
    %
    On the interior of $\{x\in\Xcal \mid V(x)=V_C(x)\}$, $V = V_C$ is a \ac{cpa} function and the proof follows similarly to \thmref \ref{thm:dissNoCA}. 
    On the interior of $\{x\in\Xcal \mid V(x)=V_P(x)\}$, $V = V_P$ is quadratic, so \thmref \ref{thm:originLMI} must be leveraged to ensure that the dissipation condition holds on $B_\epsilon(0)$. Lastly, we will establish continuity of $V$ so that its Dini Derivative is well-defined on $\Xcal$, meaning $\eqref{eq:dissInequal}$ inherits satisfaction on the boundary boundary where $V(x)=V_C(x)=V_P(x)$.

    Constraint \eqref{eq:lmiControlAffine} creates an upper bound on the QSR dissipativity condition \eqref{eq:qsrLMI} for all $x \in B_{\epsilon}(0)$ using \thmref \ref{thm:originLMI}. To develop this constraint, perform a Taylor series expansion on each function within the QSR inequality about the origin using $\nabla V = x^\top\bP + \bP x$, i.e. $x^\top\bP\hat{f} + \hat{f}^\top\bP x - \hat{h}^\top\Q\hat{h} + (x^\top\bP\hat{G} - \hat{h}^\top \Q \hat{J} - \hat{h}^\top\bS)\hat{\R}^{-1}(\hat{G}^\top\bP x - \hat{J}^\top\Q\hat{h} - \bS^\top\hat{h}) \leq 0,$ where $\hat{\R} = \R + \hat{J}^\top\bS + \bS^\top \hat{J} + \hat{h}^\top \Q \hat{h}.$ For example, 
    \begin{flalign*}
        \hat{G} = & \B + \tilde{G}(x) = \B + \bmat{\J_{g_1}(0)x & \hdots \J_{g_m}(0)x} + \\&\frac{1}{2} \bmat{\bmat{x^\top\bH_{g^{(1)}_1}(\z_{g^{(1)}_1}) \\ \vdots\\ x^\top\bH_{g^{(n)}_1}(\z_{g^{(n)}_1})}x & \hdots & \bmat{x^\top\bH_{g^{(1)}_m}(\z_{g^{(1)}_m}) \\ \vdots\\ x^\top\bH_{g^{(n)}_m}(\z_{g^{(n)}_m})}x},
    \end{flalign*}
    where the Hessian of each element is evaluated at a convex combination of $x$ and $0$.

    Factor $\x$ out on either side of the inequality. Then a Schur complement is performed about terms involving $\hat{\R}$ and another Schur complement is performed about terms involving $\Q.$ Apply the process of \thmref \ref{thm:originLMI}. Note that any term on the diagonal of the matrix that contains $x$ is bounded above using the matrix two norm, e.g. $-\bS^\top \hat{J} \leq \norm{\bS}_2\norm{\hat{J}}_2.$ The resulting \ac{lmi} is
    \begin{flalign*}
        & \bmat{\xi_1 +\bP\J_f + \J_f^\top\bP & * & * \\ 
          \B^\top\bP- \bS^\top \J_h & \bar{\R}+ \xi_2  &* \\
        \J_h & \D & \Q^{-1} {+} \frac{1}{2}(\Pi_7^{-1} {+} \Pi_{9}^{-1}) {+} \Pi_8^{-1}}\!,
    \end{flalign*}
    where $\bar{\R} = -\R - \bS^\top\D -\D^\top\bS + \Pi_4^{-1} + \frac{1}{2}(\Pi_5^{-1} +\Pi_6^{-1}),$
    \begin{flalign*}
        \xi_1{ =} & \Bigl(l_p^2\norm{\Pi_4}_2\epsilon^2\sum_{k=1}^m\norm{\J_{g_k}(0)}_2^2 {+} l_pn^{\frac{3}{2}}\beta_\epsilon \epsilon \\  
        &{ +}\frac{1}{2}(\hat{s}^2\epsilon^2n^2p\rho_\epsilon^2\norm{\Pi_5}_2 {+} l_p^2mn^3\epsilon^4\mu_\epsilon^2\norm{\Pi_6}_2 \\ & {+} \epsilon^2n^2p\rho_\epsilon^2\norm{\Pi_7}_2)\Bigr)\I,\\
         \text{ and }&\\
        \xi_2 {=} & \Bigl(\norm{\Pi_8}_2\epsilon^2\sum_{k=1}^m\norm{\J_{j_k}(0)}_2^2 + \frac{1}{2}pmn^2\theta_\epsilon^2\epsilon^4\norm{\Pi_{9}}_2 \\ & + 2\hat{s} (\epsilon^2\sum_{k=1}^m\norm{\J_{j_k}(0)}_2^2)^{\frac{1}{2}} + \hat{s}m^{\frac{1}{2}}p^{\frac{1}{2}}n\epsilon^2\theta_\epsilon\Bigr)\I. 
    \end{flalign*}
    
    Perform a Schur complement about each term in $\xi_1$, except for the term $l_pn^{\frac{3}{2}}\beta_\epsilon \epsilon$, and each term in $\xi_2$, except $\hat{s}m^{\frac{1}{2}}p^{\frac{1}{2}}n\epsilon^2\theta_\epsilon\I$, to get \eqref{eq:lmiControlAffine}.

    From \thmref \ref{thm:originLMI}, \eqref{eq:qsrLMI} holds for all $x \in B_{\epsilon}(0)$ with $V_P$ for $\Q,$ $\bS$, and $\R.$ 

    Finally, it must be shown that \eqref{eq:qsrLMI} holds for all $x \in \Xcal^{\circ}$ for the function $V.$ The function $V$ is first shown to be continuous. By \eqref{eq:Vc_higher} and the linearity of $V_C$ on each simplex, $V_C(x) \geq V_P(x)$ for all $x \in \delta \Psi.$ By \eqref{eq:Vc_lower} and the linearity of $V_C$ on each simplex, there is some region, $\Phi\subseteq B_\epsilon(0)\setminus \Psi$ and $\Psi \cap \Phi = \emptyset$, where $V_C(x) \leq V_P(x)$ for all $x \in \Phi.$ By the continuity of $V_C$ across $\Xcal\setminus\Psi^\circ$ and the Intermediate Value Theorem \cite[Theorem 3.11]{fitzpatrick2009advanced}, there must exist connected set of points $x \in \Upsilon \subseteq B_\epsilon(0)\setminus \Phi$ where $V_P(x) = V_C(x).$ Therefore, for all $y\in\Upsilon$, $\lim_{x\rightarrow y} V(x) = V(y)$; thus, $V$ is continuous across $\Xcal$  
     and \eqref{eq:qsrLMI} holds for all $x \in \Xcal^{\circ}$ for for $\Q,$ $\bS$, and $\R.$ 
    \qed

\end{pf}

The function, $V$, found in \thmref \ref{thm:dissNoCA} or \thmref \ref{thm:dissCA} is defined on a subset of the state space. In order for the dynamical system, \eqref{eq:dynSys}, to be dissipative, it must be shown that the system remains within the region.

\begin{thm}
    For dynamics \eqref{eq:dynSys} and set $\Omega \subseteq \Xcal \in \mathfrak{R}^n$, let $\Ucal_\Omega \subseteq \Ell_2^{m}$ contain a neighborhood of the origin. If \thmref \ref{thm:dissNoCA} or \thmref \ref{thm:dissCA} holds over $\Xcal,$ then the family of mappings $\Gcal_\Omega$ is locally QSR dissipative in $\Omega.$
\end{thm}
\begin{pf}
    By assumption, the QSR condition \eqref{eq:qsrLMI} holds for all $x \in \Xcal$. Thus, \eqref{eq:qsrLMI} also holds for all $x \in \Omega$ and $\bu \in \Ucal_\Omega$ and the mappings $\Gcal_\Omega(x_0)$ are locally dissipative for all $x_0\in\Omega$.

\end{pf}

\subsection{Analysis via Optimization}\label{sec:opt}

The conditions for a \ac{cpa} or combined quadratic and \ac{cpa} function to adhere to \eqref{eq:qsrLMI} developed in Theorems \ref{thm:dissNoCA} and \ref{thm:dissCA}, respectively, provide the foundation to formulate optimization problems for local dissipativity analysis. \probref \ref{prob:noLCA} and \probref \ref{prob:LCA} describe a general optimization problem to determine the dissipativity of a system while simultaneously synthesizing a storage function. \probref \ref{prob:noLCA} considers systems with no linear control affine terms, while \probref \ref{prob:LCA} considers those with linear control affine terms. In both problems, the generic cost function $J(\alpha,\beta)$ is used, where $\alpha$ and $\beta$ can be determined based of the desired QSR characterization being analyzed. Common options are given in \tabref \ref{tab:QSR}. 

\begin{problem}\label{prob:noLCA}
     Consider the mapping $\Gcal_\Omega(x_0):\Ucal_\Omega \rightarrow \Ell_{2e}^p$ defined by $\y = \Gcal_\Omega(x_0) \bu$ and \eqref{eq:dynSys}. Let $\B =\0$ and $\D = \0$. Define the triangulation $\Tcal = \{\sigma_i\}_{i=1}^{m_{\Tcal}}$ over $\Omega\in\mathfrak{R}^n,$ the \ac{cpa} function $V = \{V_x\}_{x \in \mathbb{E}_{\Tcal}}\in\mathbb{R}^n$, and the variables $\mathbf{L} = \{l_i\}_{i=1}^{m_{\Tcal}} \in \mathbb{R}$, $\alpha \vecdim{},$ and $\beta \vecdim{}.$ Let $\Pi_1 \vecdim{m\times m}$ and $\Pi_2,\Pi_3 \vecdim{p\times p}$ be diagonal matrices. The convex optimization problem is defined as
     \begin{flalign}
         &\min_{\alpha, \beta, V, \mathbf{L},\Pi} J(\alpha,\beta)\nonumber \\ 
         \text{s.t.}& \nonumber \\ 
         &\text{the constraints from \tabref \ref{tab:QSR}, \eqref{eq:dissipativityOptA}, and} \nonumber \\ 
         &\Pi_i \succeq 0, \quad \forall i\in \mathbb{Z}_1^3 \nonumber.
     \end{flalign}
\end{problem}

\begin{problem}\label{prob:LCA}
    Consider the mapping $\Gcal_\Omega(x_0):\Ucal_\Omega \rightarrow \Ell_{2e}^p$ defined by $\y = \Gcal_\Omega(x_0) \bu$ in \eqref{eq:dynSys}. Consider the sets $\Omega\in\mathfrak{R}^n$ and $\Psi\subset B_\epsilon(0)$, where $\epsilon>0.$ Define $\Tcal = \{\sigma_i\}_{i=1}^{m_{\Tcal}}$ as the triangulation of $\Omega\setminus\Psi.$ Define $V$ by \eqref{eq:V_cA} and the variables $\mathbf{L} = \{l_i\}_{i=1}^{m_{\Tcal}} \in \mathbb{R}$, $\alpha \vecdim{},$ and $\beta \vecdim{}.$ Let $\Pi_1, \Pi_5,\Pi_7 \vecdim{m\times m}$ and $\Pi_2,\Pi_3,\Pi_4, \Pi_6,\Pi_8,\Pi_9, \Pi_{10}  \vecdim{p\times p}$ be diagonal matrices. The convex optimization problem is defined as
    \begin{flalign}
         &\min_{\alpha, \beta, V, \mathbf{L}, \Pi} J(\alpha,\beta)\nonumber \\ 
         \text{s.t.}&  \nonumber \\ 
         &\text{the constraints from \tabref \ref{tab:QSR}, \eqref{opt1b:dissipativty}, and} \nonumber \\ 
         &\Pi_i \succeq 0, \quad \forall i\in \mathbb{Z}_1^{10} \nonumber.
     \end{flalign}
\end{problem}
\begin{table*}
\begin{center}
\begin{tabular}{|c |c | c | c|} 
 \hline
 \ac{io} Characterization & QSR & Design Variables & Objective 
 \\
 \hline
 $\Ell_2$-gain & $\Q = -\I$, $\bS = \0$, $\R = \gamma^2\I$ & $\alpha = \gamma^2$, $\beta = 0$, $\alpha > 0$ & $J(\alpha,\beta) = \alpha$ \\ 
 \hline
 Input Strictly Passive & $\Q = \0$, $\bS = \frac{1}{2}\I$, $\R = -\nu\I$&$\alpha = \nu$, $\beta = 0$, $\alpha > 0$ & $J(\alpha,\beta) = -\alpha$ \\
 \hline
 Output Strictly Passive & $\Q = -\rho\I$, $\bS = \frac{1}{2}\I$, $\R = \0$ & $\alpha = \rho$, $\beta = 0$, $\alpha > 0$  & $J(\alpha,\beta) = -\alpha$ \\
 \hline
 Nondegenerate conic \cite{joshi2002design,bridgeman2014conic} & $\Q = -\I$, $\bS = \frac{a+b}{2}\I,$ $\R = -ab\I$ & $\alpha = \frac{a+b}{2},$ $\beta = ab$, $\beta<0$ & $J(\alpha,\beta) = \alpha^2 - 2\beta$  \\
 \hline
 Degenerate interior conic \cite{bridgeman2014conic} & $\Q =\0$, $\bS =\I$, $\R=-d\I$ & $\alpha = d$, $\alpha<0$ & $J(\alpha,\beta) = -\alpha$  \\ 
 \hline
 QSR dissipative (verification) & $\Q \preceq 0,$ any $\bS$, $\R \succeq 0$ & $\alpha = 0$, $\beta = 0$ & $J(\alpha, \beta) = 0$ \\
 \hline
 QSR dissipative (Selection) & $\Q \preceq 0,$ any $\bS$, $\R \succeq 0$ & $\Q ,\ \bS,\ \R$ & \multicolumn{1}{p{0.2\textwidth}|}{$J(\Q ,\bS,\R)$ $ = $any convex function of its variables} \\
 \hline

\end{tabular}
\end{center}
\caption{Different QSR dissipativity characteristics can be determined in \probref \ref{prob:LCA} or \probref \ref{prob:LCA} using the design variables, objective function, and additional constraints given in the above table. This updates \cite[Table 1]{hill1977stability} to include design criteria for optimization and degenerate conic systems.}
\label{tab:QSR}
\end{table*}

\section{Solution guarantees}\label{sec:solutionGuarantess}

Finally, we establish when a \ac{cpa} or quadratic storage function is assured to be found for a locally dissipative system, as well as establishing when a combined quadratic and \ac{cpa} storage function can be found. It is important to note that the QSR dissipativity holds with some amount of strictness. \thmrefs \ref{thm:dissNoCA} and \ref{thm:dissCA} use error bounds to ensure the validity of a QSR dissipativity characterization. Therefore, the solution found by Problems 1 and 2 will always be conservative at some level.

\begin{thm}\label{thm:findSol_cpa}
    Consider \probref \ref{prob:noLCA}. Let $\Gcal_\Omega$ be locally QSR dissipative on a triangulable set, $\Omega$, with $\Q {\preceq} {-}\frac{1}{2}(\Pi_2 {+} \Pi_3)^{-1}$ and ${-}\hat{\R} {\preceq} {-}\frac{1}{2}(\Pi_1^{-1} {+}\eta\I)$, where $\eta \geq \hat{s}n\sqrt{pm}\theta_ic_{i,j} +\hat{J}_{i,j}^2$ for all $i \in \mathbb{Z}_1^{m_\Tcal}$ and $j \in \mathbb{Z}_0^n.$ Let there exist a storage function $\phi\in \mathcal{C}^2$ for which \eqref{eq:qsrLMI} holds in $\Omega,$ where $\sup_{x\in\Omega}\norm{\nabla \phi}_2 \leq C$ and $\max_{i,j\in\mathbb{Z}_1^n, x \in \Omega}\absVal{\frac{\partial \phi}{\partial x^{(i)} \partial x^{(j)}}\Big |_{x}} \leq A$ for some $C,A <\infty.$ 
    Then, \probref \ref{prob:noLCA} will find a valid \ac{cpa} storage function over $\Omega$ for \eqref{eq:dynSys} using a finite number of simplices.
\end{thm}


This proof parallels \cite[Theorem 3.3]{giesl2012construction}, which determines when a \ac{cpa} Lyapunov function can be found for an exponentially stable system, by bounded the error between the true Lyapunov function and the \ac{cpa} Lyapunov function. Here, we instead consider the QSR dissipativity condition \eqref{eq:qsrLMI} rather than Lyapunov stability conditions -- bounding the error between the system's true storage function and the \ac{cpa} storage function. This requires consideration of a matrix inequality rather than a scalar inequality.

\begin{pf}
    For some triangulation, $\Tcal$, of $\Omega$, let $\bar{t}$ be the maximum length of any simplex in $\Tcal.$ On each simplex $\sigma_i$, define
    \begin{flalign*}
        \bar{\phi}_i = \bmat{\phi(x_{i,1}) - \phi(x_{i,0}) \\ \vdots \\ \phi(x_{i,n}) - \phi(x_{i,0})}, 
    \end{flalign*}
    and define $\X_i$ as in \lemref \ref{lem:lemma_gradW}. Define also a candidate \ac{cpa} storage function over $\Tcal$ as $V(x_{i,j}) = \phi(x_{i,j})$ for all $i \in \mathbb{Z}_1^{m_\Tcal},$ $j \in \mathbb{Z}_0^n.$ By \lemref \ref{lem:lemma_gradW}, the gradient of the \ac{cpa} function $\nabla V_i$ on each simplex is $\X_i^{-1}\bar{\phi}.$

    We first aim to bound the distance between $\nabla V_i$, a linear approximation of $\nabla \phi$ across $\sigma_i,$ and the true $\nabla \phi$ evaluated at any $x \in\sigma_i$, i.e. $\norm{\X_i^{-1}\bar{\phi} -\nabla\phi(x)}_2$.
    By the properties of matrix operator norms, $\norm{\X_i^{-1}\bar{\phi} - \nabla\phi(x)}_2\leq \norm{\X_i^{-1}}_2\norm{\bar{\phi} - \X_i\nabla\phi(x)}_2.$
    
    First, let $\overline{\X} \geq \norm{\X_i^{-1}}_2$ for all $i \in \mathbb{Z}_1^{m_{\Tcal}}.$ Recall from step 3 of the proof of \cite[Theorem 5]{giesl2014revised}, $\norm{\X_i^{-1}}_2 \leq \frac{1}{\sqrt{\alpha_i}},$ where $\alpha_i$ is the minimum eigenvalue of $\X_i^\top\X_i$. Thus, $\overline{\X}$ exists and is finite.

    To bound $\norm{\bar{\phi} - \X\nabla \phi }_2,$ perform Taylor series expansion about $x \in\sigma_i$ for each case that $\phi$ is evaluated in $\bar{\phi}$ to get
    \begin{flalign*}
        \norm{\bar{\phi} - \X\nabla \phi }_2 =& \frac{1}{2}\norm{\bmat{\overline{\bH}_\phi(x_{i,1},x) - \overline{\bH}_\phi(x_{i,0},x) \\ \vdots \\ \overline{\bH}_\phi(x_{i,n},x) - \overline{\bH}_\phi(x_{i,0},x)}}_2 \\ = & \frac{1}{2}\sqrt{\sum_{j=1}^n \absVal{\overline{\bH}_\phi(x_{i,j},x) - \overline{\bH}_\phi(x_{i,0}),x}^2}.
    \end{flalign*}
    By the triangle inequality and the properties of matrix operator norms, this is bounded above by 
    \begin{flalign*}
        & \frac{1}{2}\sqrt{\sum_{j=1}^n (\norm{\bH_\phi(z_{i,1}\!)}_2\!\norm{x_{i,1}{-}x}_2^2 \!{+}\!\norm{\bH_\phi(z_{i,0}\!)}_2\!\norm{x_{i,0}{-}x}_2^2)^2}
    \end{flalign*}
    By \cite[Lemma 2.3]{giesl2012construction}, this is bounded above by $An^{\frac{3}{2}}\bar{t}^2.$ Thus, $\norm{\X_i^{-1}\bar{\phi} - \nabla\phi(x)}_2$\\ $\leq \norm{\X_i^{-1}}_2\norm{\bar{\phi} - \X_i\nabla\phi(x)}_2 \leq \overline{\X}An^{\frac{3}{2}}\bar{t}^2.$

    Consider the matrix $\M$ in Constraint \eqref{eq:constraint_QSRLMI_noCA}. For any element of $\M$ that contains $\nabla V$, add and subtract $\nabla \phi,$ e.g. $\nabla V^\top f(x_{i,j}) = (\nabla V - \nabla \phi +\nabla \phi)^\top f(x_{i,j}).$ For each simplex, $\nabla V = \X_i^{-1}\bar{\phi}.$ Then, by the Schur complement \cite{boyd1994linear}, Constraint \eqref{eq:constraint_QSRLMI_noCA} is equivalent to the inequality,
    \begin{flalign*}
        & \Lambda + (\X^{-1}\bar{\phi} {-} \nabla \phi)^\top f(x_{i,j})\\& + \frac{1}{2}(\X^{-1}\bar{\phi} {-} \nabla \phi)^\top G(x_{i,j})\tilde{\R}^{-1}\\ &\left(G(x_{i,j})^\top(\X^{-1}\bar{\phi} {-} \nabla \phi) {-} \bS^\top h(x_{i,j}) {-} J(x_{i,j})^\top\tilde{\Q}h(x_{i,j})\right) \\ & {+} \hat{f}_{i,j} {+} \frac{1}{2\underline{\pi}_1}\hat{s}^2\hat{h}_{i,j}^\top\hat{h}_{i,j} {+} \frac{1}{4\underline{\pi}_1}l_i^2\hat{G}_{i,j}^\top \hat{G}_{i,j} {+} \frac{1}{2}\hat{h}_{i,j}^\top\Pi_2 h_{i,j} \leq 0,
    \end{flalign*}
    where
    \begin{flalign*}
        \Lambda \!\!= & \nabla \phi^\top f(x_{i,j}) {-} h(x_{i,j})^\top\tilde{\Q} h(x_{i,j}) \\ &{+} (\frac{1}{2}\nabla \phi^\top G(x_{i,j}) {-} h(x_{i,j})^\top \bS {-}h(x_{i,j})^\top \tilde{\Q} J(x_{i,j}))\\ &\tilde{\R}^{-1}( \frac{1}{2}G(x_{i,j})^\top \nabla \phi {-} \bS^\top \!h(x_{i,j}) {-}J(x_{i,j})\!^\top\! \tilde{\Q} h(x_{i,j})).
    \end{flalign*}
    Evaluating \eqref{eq:qsrLMI} at $x_{i,j}$ using the asserted storage function, $\phi,$ equals $\Lambda$. Further,
    $\tilde{\R} = \hat{\R} + \frac{1}{2}(\Pi_1^{-1} + \hat{s}\sqrt{pm}\theta_i c_{i,j}\I),$ and
    $\tilde{\Q} = (\Q^{-1} + \frac{1}{2}\Pi_2^{-1} + \frac{1}{2}\Pi_3^{-1})^{-1},$ and $\hat{f}_{i,j},$ $\hat{h}_{i,j}$, $\hat{G}_{i,j}$, and $\hat{s}$ are defined by \thmref \ref{thm:dissNoCA}.

    Using the properties of matrix operator norms, the inequality is bounded above by
    \begin{flalign*}
        & \Lambda {+}\Bigl(\!\norm{f(x_{i,j})}_2 \!{+} \norm{\X_i^{-1}}_2\!\norm{\bar{\phi} {-} \X\nabla \phi}_2\!\norm{\tilde{\R}^{-1}}_2\!\norm{G(x_{i,j})}_2 \\ & {+} \norm{\tilde{\R}^{-1}}_2\norm{h(x_{i,j})}_2(\norm{\bS} {+}\norm{\tilde{\Q}}_2\norm{J(x_{i,j}}_2)\!\Bigr)\\&\Bigl(\norm{\X_i^{-1}}_2\norm{\bar{\phi} - \X\nabla \phi}_2\Bigr)+ \hat{f}_{i,j} + \frac{1}{2\underline{\pi}_1}\hat{s}^2\hat{h}_{i,j}^\top\hat{h}_{i,j} \\ &+ \frac{1}{4\underline{\pi}_1}l_i^2\hat{G}_{i,j}^\top \hat{G}_{i,j} + \frac{1}{2}\hat{h}_{i,j}^\top\Pi_2 \hat{h}_{i,j}.
    \end{flalign*}
    The terms $\hat{f}_{i,j},$ $\hat{h}_{i,j},$ and $\hat{G}_{i,j}$ all contain $c_{i,j}$ defined either by \eqref{eq:oldC} if $0 \in \sigma_i$ or by \eqref{eq:c} if $0 \notin \sigma.$ By definition, each $c_{i,j} \leq 2\bar{t}^2.$ Altogether, the following inequality upper bounds Constraint \eqref{eq:constraint_QSRLMI_noCA}:
    \begin{flalign*}
        & \Bigl(\norm{f(x_{i,j})}_2 {+} \bar{\X}An^{\frac{3}{2}}\bar{t}^2\norm{\tilde{\R}^{-1}}_2\norm{G(x_{i,j}}_2 \\ & + \norm{\tilde{\R}^{-1}}_2\norm{h(x_{i,j})}_2(\norm{\bS} +\norm{\tilde{\Q}}_2\norm{J(x_{i,j}}_2)\Bigr)\bar{\X}An^{\frac{3}{2}}\bar{t}^2 \\ &+ nl_i\beta_i\bar{t}^2 + \frac{2}{\underline{\pi}_1}\hat{s}^2n^2\bar{t}^4\sum_{k=1}^p\rho_{k}^2  + \frac{1}{\underline{\pi}_1}l_i^2n^2\bar{t}^4\sum_{k=1}^m\mu_k^2 \\ &+ 2 n^2\bar{t}^4\sum_{k=1}^p \rho_k^2\pi_{2,k} \leq -\Lambda .
    \end{flalign*}

    By assumption $\Lambda < 0$ at $x \neq 0$ and $\Lambda = 0$ at $x = 0$, thus $-\Lambda \geq 0$ for all $x \in \Omega.$ If the above inequality holds for all $x_{i,j}\neq 0$, then $V$ is a viable storage function. Each term on the left hand side of the inequality contains a term $\bar{t}$, while the term on the right hand side is constant at any $x \in \Omega$. Therefore, $\bar{t}$ approaches $0$, the above inequality will hold for any $x_{i,j} \neq 0.$ Therefore, \probref \ref{prob:noLCA} will have a solution. \qed
\end{pf}


Next, we establish conditions for which a quadratic storage function will be found in a ball about the origin for a locally dissipative system.

\begin{thm}\label{thm:findSol_quad}
    Consider \probref \ref{prob:LCA}. Let $\Gcal_\Omega$ be locally QSR dissipative on $\Omega$ with $\Q \preceq -\frac{1}{2}(\Pi_7 + \Pi_8 + 2\Pi_9)^{-1}$ and $-\hat{\R} \preceq -\frac{1}{2} (2\Pi_4^{-1} + \Pi_5^{-1} +\Pi_6^{-1})$, where $\eta \geq \hat{s}n\sqrt{pm}\theta_ic_{i,j}+\hat{J}_{i,j}^2$ for all $i \in \mathbb{Z}_1^{m_\Tcal}$ and $j \in \mathbb{Z}_0^n.$ Let there exist a storage function $\Phi(x) = x^\top\bP x$ $(\bP \succ 0)$ such that \eqref{eq:qsrLMI} holds for \eqref{eq:dynSys} when linearized about the origin.
    Then, \probref \ref{prob:LCA} will find a valid storage function, $V_P$ over $B_\epsilon(0)$ for \eqref{eq:dynSys} using a non-zero $\epsilon$.
\end{thm}
\begin{pf}
    Consider the region $B_\epsilon(0).$ In \probref \ref{prob:LCA}, Constraint \eqref{eq:lmiControlAffine} is equivalently expressed as
    \begin{flalign*}
        x^\top&\Bigl( \J_f^\top \bP + \bP \J_f + \omega_1 - \J_h^\top \Q^{-1}\J_h \\ & -(\bP\B - \J_h^\top \Q \D - \J_h^\top \bS)\bar{\R}^{-1}\\&(\B^\top \bP-\D^\top \Q \J_h - \bS^\top \J_h)\Bigr)x \\& \leq -x^\top \omega_1 x, 
    \end{flalign*}
    where $\omega_1=(l_p^2\norm{\Pi_5}_2\epsilon^2\sum_{k=1}^m\norm{\J_{g_k}(0)}_2^2 + l_pn^{\frac{3}{2}}\beta_\epsilon \epsilon +\frac{1}{2}(\hat{s}^2\epsilon^2n^2p\rho_\epsilon^2\norm{\Pi_6}_2 + l_p^2mn^3\epsilon^4\mu_\epsilon^2\norm{\Pi_7}_2 + \epsilon^2n^2p\rho_\epsilon^2\norm{\Pi_8}_2))\I$, $\bar{\R} = -\R -\bS^\top\D -\D^\top\bS -\D^\top\Q\D + \Pi_5^{-1}  + \frac{1}{2}(\Pi_6 + \Pi_7) + (\norm{\Pi_9}_2\epsilon^2\sum_{k=1}^m\norm{\J_{j_k}(0)}_2^2 + \frac{1}{2}pmn^2\theta_\epsilon^2\epsilon^4\norm{\Pi_{10}}_2 + 2\hat{s} \sqrt{\epsilon^2\sum_{k=1}^m\norm{\J_{j_k}(0)}_2^2} + \hat{s}\sqrt{mp}n\epsilon^2\theta_\epsilon)\I.$ 
    
    An upper bound on this constraint can be found when $\bar{\R}$ is replaced with $\hat{\R} = -\R -\bS^\top\D -\D^\top\bS -\D^\top\Q\D$. Further, if
    \begin{flalign*}
        x^\top& \Bigl( \J_f^\top \bP + \bP \J_f - \J_h^\top \Q^{-1}\J_h \\& +(\J_h^\top \Q \D + \J_h^\top \bS - \bP\B)\hat{\R}^{-1}\\&(\D^\top \Q \J_h + \bS^\top \J_h - \B \bP)\Bigr)x \\ & \leq -m\omega_1\epsilon^2\leq -x^\top \omega_1 x
    \end{flalign*}
    holds, then so does Constraint \eqref{eq:lmiControlAffine}.
    

    Because there is an $\epsilon$ in each term of $\omega_1$, there is some $\epsilon$ small enough that the above inequality holds. For this $B_\epsilon(0)$, $\bP$ will be a viable solution for the quadratic storage function. \qed



\end{pf}



Finally, we establish when a \ac{cpa} and quadratic storage function can be determined for a locally dissipative system.

\begin{thm}\label{thm:findSol_quadAndCPA}
    Consider \probref \ref{prob:LCA}. Let $\Gcal_\Omega$ be locally QSR dissipative on $\Omega$ with $\Q \preceq \frac{1}{2}\min (-(\Pi_2 + \Pi_3)^{-1}, -(\Pi_7 + \Pi_8 + 2\Pi_9)^{-1}$ and $-\hat{\R} \preceq \frac{1}{2}\min(-(\Pi_1^{-1} +\eta\I), -(2\Pi_4^{-1} + \Pi_5^{-1} +\Pi_6^{-1})$, where $\eta \geq \hat{s}n\sqrt{pm}\theta_ic_{i,j}+\hat{J}_{i,j}^2$ for all $i \in \mathbb{Z}_1^{m_\Tcal}$ and $j \in \mathbb{Z}_0^n.$ Let there exist a storage function $\phi\in \mathcal{C}^2$ for which \eqref{eq:qsrLMI} holds in $\Omega,$ where $\sup_{x\in\Omega}\norm{\nabla \phi}_2 \leq C$ and $\max_{i,j\in\mathbb{Z}_1^n, x \in \Omega}\absVal{\frac{\partial \phi}{\partial x^{(i)} \partial x^{(j)}}\Big |_{x}} \leq A$ for some $C,A <\infty,$ and let there exist a Lyapunov function $\Phi(x) = x^\top\bP x$ $(\bP \succ 0)$ such that \eqref{eq:qsrLMI} holds for \eqref{eq:dynSys} when linearized about the origin and $\Phi(x)\geq \phi(x)$ for all $x \in B_\epsilon(0)$.
    Then, \probref \ref{prob:LCA} will find a valid storage function, $V$ over $\Omega$ for \eqref{eq:dynSys} using a non-zero $\epsilon$ and a finite number of simplices.
\end{thm}

\begin{pf}

    On $B_\epsilon(0)$ the theorem holds due to \thmref \ref{thm:findSol_quad}. Next, consider $\Omega$ as a whole. Like step 3 of \cite[Theorem 3.3]{giesl2012construction}, we construct a continuous storage function by considering the convex combination of two storage functions.
    Define $\rho:B_\epsilon\setminus \Psi\rightarrow\IR_0^1$ as a non-decreasing, continuous function where $\rho(x) = 0$ for all $x\in \delta\Psi$ and $\rho(x)=1$ for all $x \in \delta (\Omega\setminus B_\epsilon (0)).$ Define the function 
    \begin{flalign*}
    V_\phi(x) = 
        \begin{cases}
            \Phi(x) & x \in \Psi \\
            \rho(x)\Phi(x) + (1-\rho(x))\phi(x) & x \in B_\epsilon (0)\setminus \Psi \\
            \phi(x) & x \in \Omega\setminus B_\epsilon (0).
        \end{cases}
    \end{flalign*}
    By assumption, both $\Phi$ and $\phi$ are storage functions. The function $V_\phi$ is continuous by the composition of continuous functions. Further, $V_\phi$ a storage function as by \cite[Theorem 3]{willems1972dissipative} the convex combination of storage functions is also a storage function. Note also that $\sup_{x \in \Omega} \norm{\nabla V_\phi}_2 \leq C_\phi$ and $\max_{i,j\in\mathbb{Z}_1^n, x \in \Omega}\absVal{\frac{\partial \phi}{\partial x^{(i)} \partial x^{(j)}}\Big |_{x}} \leq A_\phi$ for some $C_\phi,A_\phi <\infty.$ Thus, by \thmref \ref{thm:findSol_cpa}, a viable \ac{cpa} storage function will be found across $\Omega\setminus B_\epsilon(0)$ that adheres to $V_\phi(x).$ This function will also adhere to \eqref{eq:Vc_higher} and \eqref{eq:Vc_lower} by definition of $V_\phi.$ Thus, \probref \ref{prob:LCA} will is able to determine a storage function with finite simplices and a non-zero $\epsilon$ about the origin. \qed



\end{pf}

\section{Numerical examples}\label{sec:numRes}

We demonstrate the efficacy of our method on three systems. We first find local conic bounds for a one-dimensional nonlinear system. Here, we must compare to bounds found through ad-hoc ingenuity because we know of no existing, similar technique to systematically seek generic conic bounds of nonlinear systems. Next, we determine the local $\Ell_2$ gain of a two-dimensional pendulum. Finally, we compare our results to that of \cite{summers2013quantitative} for a three-dimensional polynomial system.

\subsection{Nonlinear conic system}

\begin{figure}
    \centering
    \resizebox{.95\columnwidth}{!}{%
%
%
\begin{tikzpicture}

\begin{axis}[%
width=4.521in,
height=1.378in,
at={(0.758in,2.554in)},
scale only axis,
xmode=log,
xmin=10,
xmax=100000,
xminorticks=true,
ymin=0.45,
ymax=0.55,
ylabel style={font=\bfseries\color{white!15!black}},
ylabel={b},
axis background/.style={fill=white},
title style={font=\bfseries},
title={Upper Conic Bound},
axis x line*=bottom,
axis y line*=left,
xmajorgrids,
xminorgrids,
ymajorgrids,
legend style={at={(0.565,0.806)}, anchor=south west, legend cell align=left, align=left, draw=white!15!black}
]
\addplot [color=black, line width=1.5pt, mark=*, mark options={solid, fill=black, black}]
  table[row sep=crcr]{%
20	0.531914895755377\\
200	0.500300185376784\\
2000	0.500003006446097\\
20000	0.500000075078707\\
};
\addlegendentry{Bound found via Problem 2}

\addplot [color=blue, dashed, line width=1.5pt]
  table[row sep=crcr]{%
10	0.500000038805563\\
20	0.500000038805563\\
200	0.500000038805563\\
2000	0.500000038805563\\
20000	0.500000038805563\\
100000	0.500000038805563\\
};
\addlegendentry{Analytical Bound}

\end{axis}

\begin{axis}[%
width=4.521in,
height=1.378in,
at={(0.758in,0.481in)},
scale only axis,
xmode=log,
xmin=10,
xmax=100000,
xminorticks=true,
xlabel style={font=\bfseries\color{white!15!black}},
xlabel={Number of Simplexes},
ymin=-0.05,
ymax=0.05,
ylabel style={font=\bfseries\color{white!15!black}},
ylabel={a},
axis background/.style={fill=white},
title style={font=\bfseries},
title={Lower Conic Bound},
axis x line*=bottom,
axis y line*=left,
xmajorgrids,
xminorgrids,
ymajorgrids
]
\addplot [color=black, line width=1.5pt, mark=*, mark options={solid, fill=black, black}, forget plot]
  table[row sep=crcr]{%
20	-0.0305445264731306\\
200	-0.0322703307668048\\
2000	-0.0322919478942869\\
20000	-0.032634525422286\\
};
\addplot [color=blue, dashed, line width=1.5pt, forget plot]
  table[row sep=crcr]{%
10	-5.77236312829019e-09\\
20	-5.77236312829019e-09\\
200	-5.77236312829019e-09\\
2000	-5.77236312829019e-09\\
20000	-5.77236312829019e-09\\
100000	-5.77236312829019e-09\\
};
\end{axis}

\begin{axis}[%
width=5.833in,
height=4.375in,
at={(0in,0in)},
scale only axis,
xmin=0,
xmax=1,
ymin=0,
ymax=1,
axis line style={draw=none},
ticks=none,
axis x line*=bottom,
axis y line*=left
]
\end{axis}
\end{tikzpicture}%
	}
    \caption{The analytical upper (b) and lower (a) conic bounds of \eqref{eq:nonlinearConic} are determined via \probref \ref{prob:LCA} for an increasing number of simplices. We are able to determine conic bounds of the system that closely match the analytical bounds. }
    \label{fig:conicResults}
\end{figure}
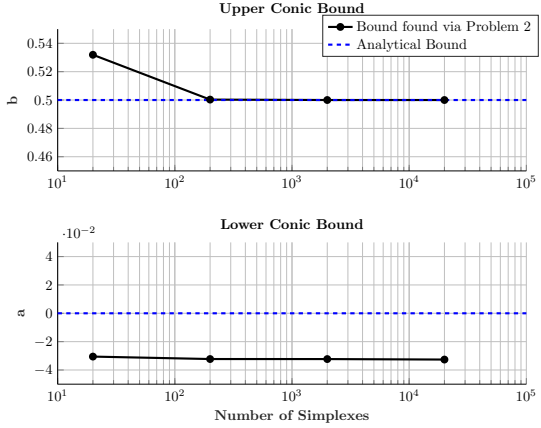

We consider the nonlinear system
\begin{flalign}\label{eq:nonlinearConic}
    \Gcal_\Omega (x_0): \begin{cases}
        \dot{x} &= k_1x^3 - (k_1+ k_2)x + Bu \\
        y &= Cx + Du,
    \end{cases}
\end{flalign}
where $k_1,k_2 > 0$. Let $\Xcal = [-1,1]$, where $\Omega = [-0.8092, 0.8092]$ is an RPI set found using the SOSAnalysis toolbox \cite{packard2009}. When $k_1 = 1$, $k_2= 2$, $B=1$, $C = 1$, and $D = 0,$ the system is locally nondegenerate conic, i.e., $\Gcal_\Omega \in \text{cone}(a, b)$, where $a = -5.8\times 10^{-9}$ and $b = 0.50$ (see \secref \ref{sec:conicBound1D}). We then implement \probref \ref{prob:LCA} to determine the conic bounds via storage function synthesis for an increasing number of simplices. Specifically, we consider a triangulation over evenly spaced simplices with $\Delta = 0.1, 0.01, 0.001, 0.0001$, where $\epsilon = 2\Delta.$ As seen in Figure \ref{fig:conicResults}, the conic bounds determined by \probref \ref{prob:LCA} closely match the analytical bounds -- even with a relatively small number of simplices.

\subsection{$\Ell_2$ gain of Pendulum}
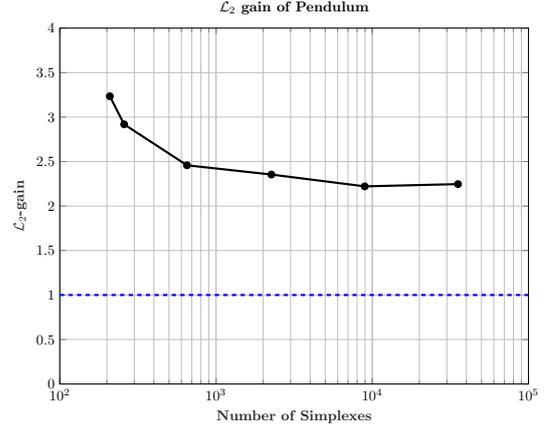
\begin{figure}
    \centering
    \resizebox{.95\columnwidth}{!}{%
%
%
\begin{tikzpicture}

\begin{axis}[%
width=4.521in,
height=3.459in,
at={(0.758in,0.551in)},
scale only axis,
xmode=log,
xmin=100,
xmax=100000,
xminorticks=true,
xlabel style={font=\bfseries\color{white!15!black}},
xlabel={Number of Simplexes},
ymin=0,
ymax=4,
ylabel style={font=\bfseries\color{white!15!black}},
ylabel={$\mathcal{L}_2$-gain},
axis background/.style={fill=white},
title style={font=\bfseries},
title={$\mathcal{L}_2$ gain of Pendulum},
xmajorgrids,
xminorgrids,
ymajorgrids
]
\addplot [color=black, line width=1.5pt, mark=*, mark options={solid, fill=black, black}, forget plot]
  table[row sep=crcr]{%
209	3.23369611847569\\
258	2.91848756709094\\
652	2.45796182726605\\
2260	2.35400618921019\\
8954	2.22124220886908\\
35353	2.24658619778887\\
};
\addplot [color=blue, dashed, line width=1.5pt, forget plot]
  table[row sep=crcr]{%
100	1\\
209	1\\
258	1\\
652	1\\
2260	1\\
8954	1\\
35353	1\\
100000	1\\
};
\end{axis}

\begin{axis}[%
width=5.833in,
height=4.375in,
at={(0in,0in)},
scale only axis,
xmin=0,
xmax=1,
ymin=0,
ymax=1,
axis line style={draw=none},
ticks=none,
axis x line*=bottom,
axis y line*=left
]
\end{axis}
\end{tikzpicture}%
	}
    \caption{The $\Ell_2$ gain of the pendulum found on a triangulation over $\Xcal \supseteq \Omega$ for an increasing number of simplices. As the triangulation becomes more fine, the gain found by \probref \ref{prob:LCA} becomes closer to the analytical bound.}
    \label{fig:pendulumResults}
\end{figure}

\begin{figure}
    \centering
    \includegraphics[width=0.95\linewidth]{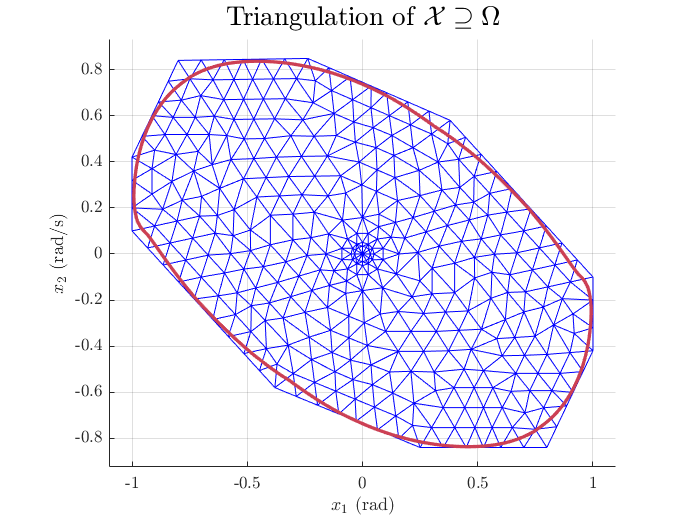}
    \caption{An example of a triangulation over $\Xcal\supseteq \Omega$, where $\Omega$ is an RPI set denoted by the red line. The set $\Omega$ was found using \cite[Theorem 5]{lavaei2025L2} and is valid for inputs where $\absVal{u}< 0.1942.$ }
    \label{fig:pendBF}
\end{figure}

\begin{figure}
    \centering
    \includegraphics[width=0.95\linewidth]{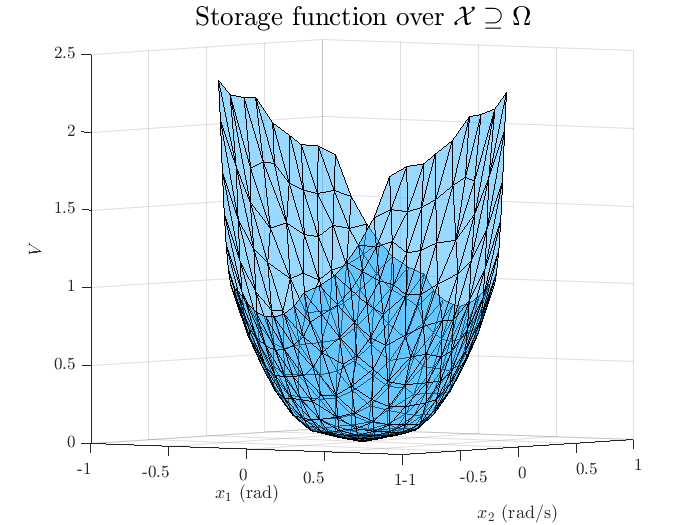}
    \caption{An example of a storage function over $\Xcal \supseteq \Omega$ found by \probref \ref{prob:LCA}.}
    \label{fig:pendV}
\end{figure}

Consider the dynamics of a pendulum,
\begin{equation}
    \Gcal_\Omega (x_0): \begin{cases}
        \dot{x}_1 & = x_2 \\
        \dot{x}_2 & = -\sin x_1 -x_2 + u \\
        y & = x_2,
    \end{cases}
\end{equation}
where $x \vecdim{n}$, $u \vecdim{}$, and $y\vecdim{}$. Let $\Xcal \supseteq\Omega$ be the triangulated region about the origin, shown in Figure \ref{fig:pendBF}, where the RPI set $\Omega$ is found using \cite[Theorem 5]{lavaei2025L2}, where the system is able to have inputs of $\absVal{u}< 0.1942$. \probref \ref{prob:LCA} was applied to the system -- solving for the $\Ell_2$ gain specifically. Because the pendulum has a sinusoidal term, \ac{sos} techniques \cite{summers2013quantitative} cannot be used for analysis. Instead, we compare to the gain found using the global analytical Lyapunov function $V(x) = (1-\cos x_1) + \frac{1}{2}x_2^2$, which results in the analytical gain bound $\gamma \geq 1.$ Figure \ref{fig:pendulumResults} shows the gain found on $\Omega$ for an increasing number of simplices, where the best result is $\gamma = 2.221$ Figure \ref{fig:pendV} shows an example of a storage function found across $\Omega$ with 652 simplices.

\subsection{Three dimensional polynomial system}

Finally, we implemented \probref \ref{prob:LCA} to determine the $\Ell_2$ gain of the following three-dimensional polynomial system,
\begin{flalign}\label{eq:poly3d}
\Gcal_\Omega(x_0): \begin{cases}
    \dot{x}_1 & = -x_1-x_3 +x_2 -x_3x_2^2 \\
    \dot{x}_2 & = -x_2x_3^2 -x_2 + u \\
    \dot{x}_3 & = \frac{1}{2}(x_1 - x_3) \\
    y & = x_2
\end{cases}
\end{flalign}
over $\Xcal  = [-0.5, 0.5]\times [-0.5, 0.5] \times [-0.5, 0.5],$ where $\Omega \subseteq \Xcal$ is a RPI set determined using the SOSAnalysis package \cite{packard2009}. Using 24,576 simplices, the gain was determined to be $\gamma \geq 1.65.$ In contrast, when applying \cite{packard2009}, a viable gain was searched for using the SOSAnalysis package using a simple line search. The gain found by \cite{packard2009} was $\gamma \geq 1.01$.

\section{Conclusion}\label{sec:Concl}
This paper presents a general method to determine the local dissipativity properties of nonlinear control affine systems. We develop a convex optimization program that simultaneously optimizes the desired \ac{io} characterization of the system and synthesizes the corresponding storage function. Further, we provide guarantees that our analysis method will always be feasible when determining \ac{cpa} or quadratic storage functions for systems with strict local dissipativity properties. Prior results establishing dissipativity characterizations are either limited to polynomial systems, only consider gain, or rely on ad hoc, ingenuity. Numerical results show that our techniques produce similarly tight characterizations to alternatives (when those alternatives are available).


\appendix
\section{Function Error Bounds}\label{sec:fxnError}

\lemref \ref{lem:fxnBound} determines a bound on the behavior of a function across a simplex.

\begin{lem}\label{lem:fxnBound}
    Consider $\zeta: \map{n}{m},$ where $\zeta \in \mathcal{C}^2$. Let $\sigma \defeq \co\{x_j\}_{j=0}^n$ be an n-simplex in $\mathbb{R}^n.$ If $\sum_{j=0}^n\lambda_jx_j \in \sigma$, then
     \begin{flalign}\label{eq:fxnBound}
        \norm{\zeta(x) - \sum_{j=0}^n \lambda_j \zeta(x_j)}_2 &= \frac{1}{2}\sumLam\norm{\overline{\bH}_{\zeta}(x_j,x)}_2 \\ &\leq \frac{1}{2}n\sumLam c_j (\sum_{k=0}^m\mu_k) \nonumber \text{ and }\\
        \label{eq:fxnBound_inf}
        \norm{\zeta(x) - \sum_{j=0}^n \lambda_j \zeta(x_j)}_\infty &= \frac{1}{2}\sumLam\norm{\overline{\bH}_{\zeta}(x_j,x)}_\infty \\ &\leq \frac{1}{2}n\max_{k\in\mathbb{Z}_1^m}\mu_k\sumLam   c_j, \nonumber
        \end{flalign}
        where
        \begin{flalign}
        \label{eq:c}
        & c_j = \max_{\nu \in \mathbb{Z}_0^n}\norm{x_j - x_{\nu}}_2^2 \text{ and }\\
        & \mu_k \geq \max_{q,r \in \mathbb{Z}_1^n, \xi \in\sigma} \absVal{ \frac{\partial^2 \zeta ^{(k)}(x)}{\partial x ^{(q)}\partial x^{(r)}}\Big\vert_{x = \xi}}.\label{eq:muZeta}
    \end{flalign}
\end{lem}
\begin{pf}
    By definition, any point $x \in \sigma$ can be written as a convex combination of the vertices, i.e., $x = \sum_{j=0}^n \lambda_j x_j.$ Let $\tilde{e}(x) = \zeta(x) - \sum_{j=0}^n \lambda_j \zeta(x_j).$ Consider a dimension of $\zeta$, represented as $\zeta^{(k)}$, and apply \thmref \ref{thm:taylor} about $x$ for each vertex point in $\tilde{e}(x)$ to get
    \begin{flalign*}
        \tilde{e}(x) = \zeta(x) - \sum_{j=0}^n \lambda_j &\Big( \zeta^{(k)}(x) + \langle \nabla \zeta^{(k)}(x), \Delta x_j\rangle  \\ & + \frac{1}{2}\overline{\bH}_{\zeta^{(k)}}(x_{j},x)\Big).
    \end{flalign*}
    
    In each dimension, the zero-th order term of each function's Taylor expansion cancel with $\zeta^{(k)}(x)$ and the summed first order terms are zero as $\sum_{j=0}^n \lambda_j \Delta x_j= \sum_{j=0}^n \lambda_j x_j - x = 0.$ Altogether,
    \begin{flalign*}
        \tilde{e}(x) = -\frac{1}{2}\sum_{j=0}^n \lambda_j\bmat{ \overline{\bH}_{\zeta^{(1)}}(x_{j},x) \\ \vdots \\ \overline{\bH}_{\zeta^{(m)}}(x_{j},x) }  = -\frac{1}{2}\sum_{j=0}^n \lambda_j\overline{\bH}_\zeta(x_j,x).
    \end{flalign*}
    To find \eqref{eq:fxnBound}, recall that $\norm{\tilde{e}(x)}_2 \leq \norm{\tilde{e}(x)}_1.$ Therefore,
    \begin{flalign*}
        \norm{\tilde{e}(x)}_2 \leq \frac{1}{2}\sum_{j=0}^n \lambda_j\sum_{k=1}^m \absVal{\overline{\bH}_{\zeta^{(k)}}(x_{j},x)}.
    \end{flalign*}
     Through application of the Cauchy Schwarz inequality and the properties of matrix operator norms, 
     \begin{flalign*}
         \sum_{k=1}^m \absVal{\overline{\bH}_{\zeta^{(k)}}(x_{j},x)} \leq \sum_{j=0}^n \lambda_j \sum_{k=1}^m\norm{\bH_{\zeta^{(k)}}(z)}_2\norm{x_j-x}_2^2.
     \end{flalign*}
     Then, \cite[Lemma 2.3]{giesl2012construction}, and the bound $\norm{x_j-x}_2^2 \leq \max_{\nu \in \mathbb{Z}_0^n}\norm{x_j - x_\nu}_2^2$ are applied to find \eqref{eq:fxnBound}. 

      To find \eqref{eq:fxnBound_inf}, instead consider
      \begin{flalign*}
          \norm{\tilde{e}(x)}_{\infty} = \frac{1}{2}\sum_{j=0}^n \lambda_j\max_{k\in \mathbb{Z}_1^m}\absVal{\overline{\bH}_{\zeta^{(k)}}(x_{j},x)}
      \end{flalign*}
      The same procedure as above is applied to get \eqref{eq:fxnBound_inf}.
     \qed
\end{pf}

\section{LMI Bound}\label{sec:lmiBoundOrigin}

\begin{cor}\label{cor:lmi_bound_gh}
    Consider \eqref{eq:genericMat} as defined in \thmref \ref{thm:lmi_bound}. Let $\sigma \defeq \co\{x_j\}_{j=0}^n$ be an n-simplex in $\mathbb{R}^n.$ If $x = \sum_{j=0}^n\lambda_jx_j \in \sigma,$ then
    \begin{flalign}\label{eq:lmi_boundSchur_2}
        \M(x) - &\sum_{j=0}^n \lambda_j \M(x_j) \preceq  \nonumber \\ &
        \sum_{j=0}^n \lambda_j \bmat{\frac{1}{2}\tilde{\phi}_j & *  & * \\ \0 & \frac{1}{2}\Pi^{-1} & * \\ \tilde{\zeta} & \0 & -2\Pi^{-1}} \defeqRight \E(x),
    \end{flalign}
    where $\tilde{\phi}_j \geq \max_{j\in\mathbb{Z}_0^n} \absVal{\overline{\bH}_\phi(x,x_0) - \overline{\bH}_\phi(x_j,x_0)}$, $\hat{\zeta} \vecdim{m\times 1}$ where each element \\ $\tilde{\zeta}^{(k)} \geq \max_{j\in\mathbb{Z}_0^n}\absVal{\overline{\bH}_{\zeta^{(k)}}(x,x_0)- \overline{\bH}_{\zeta^{(k)}}(x_j,x_0)}$ for $k \in \mathbb{Z}_1^m$, and 
    $\Pi\in \mathbb{R}^{m \times m}$ is a diagonal, positive definite matrix with elements $\pi_k,$ $k \in \mathbb{Z}_1^m.$
    
    Moreover, if $\M(x_j) + \E(x_j) \preceq 0$ holds for all vertex points of $\sigma,$ then $\M(x) \preceq 0$ holds for all $x \in \sigma.$
\end{cor}

This proof parallels the Taylor series expansion method of \cite[Proposition 2.2]{giesl2012construction} for each individual function within $\M(x).$ As with \lemref \ref{lem:lemma_eq_bound}, the bound developed in \corref \ref{cor:lmi_bound_gh} is essential for enforcing \acp{lmi} on simplices that contain the origin as a vertex.

\begin{pf}
    Let $\tilde{\E}(x) = \M(x) - \sumLam \M(x_j),$ where in $\M(x),$ $\phi(x) = \phi(x_0) + \nabla \phi(x_0)^\top (x-x_0) + \frac{1}{2}\overline{\bH}_\phi(x,x_0)$ and $\zeta(x) = \zeta(x_0) + \nabla \zeta(x_0)^\top (x-x_0) + \frac{1}{2}\overline{\bH}_\zeta(x,x_0)$. As in Lemma \ref{lem:fxnBound}, apply a \thmref \ref{thm:taylor} to $\phi$ and $\zeta$ about $x$ for each vertex point to get 
    \begin{flalign*}
        &\frac{1}{2}\sum\limits_{{j}=0}^n\lambda_{j}\!\!
        \bmat{{-} \overline{\bH}_{\phi}(x,x_0) {+} \overline{\bH}_{\phi}(x_j,x_0) & \0\\
        \0 & \0}\!\!\\& {+} \text{He}\!\!\left(\! \bmat{\0 & (-\overline{\bH}_\zeta(x,x_0)+\overline{\bH}_\zeta(x_j,x_0))^\top\\ \0 & \0}\!\!\bmat{\0 & \0 \\ \0& \I}\!\right)\!\!.
    \end{flalign*}
    From here, the proof follows as \thmref \ref{thm:lmi_bound}. \qed
\end{pf}

\section{Analytical Conic Bound}\label{sec:conicBound1D}

\begin{lem}
   Consider \eqref{eq:nonlinearConic} over $\Omega\subseteq \Xcal$, where $\Omega$ is a RPI set and $\Xcal= [-1,1]$. If there exists a $\bP > 0$ for which
    \begin{flalign}\label{eq:conicConditions}
        \bmat{-k_2\bP{+}C^\top C & \frac{1}{2}\bP^\top B {+}C^\top D {-} C^\top\frac{a+b}{2}\I\\ \frac{1}{2} B^\top \bP {+}D^\top C {-} \frac{a+b}{2}\I C & ab\I {+}D^\top  D {-} (a{+}b)\I D } \preceq 0,
    \end{flalign}
    then $\Gcal_\Omega \in \text{cone}(a,b)$. 
\end{lem}
\begin{pf}
Consider the storage function $V = \frac{1}{2}\bP x^2.$ For the system to be locally dissipative, 
\begin{flalign*}
    x^\top\bP \dot{x} &\leq  \\ &(Cx+Du)^\top \Q (Cx+Du) + 2(Cx+Du)^\top \bS u\\& + u^\top \R u,
\end{flalign*}
must be satisfied for all $x \in \Omega$.
The above inequality is equivalent to
\begin{flalign*}
    k_1\bP x^2 - k_1\bP x^4 & \geq\\&  -k_2\bP x^2 + x\bP B u \\&-(Cx+Du)^\top \Q (Cx+Du) \\&- 2(Cx+Du)^\top \bS u - u^\top \R u.
\end{flalign*}
Because $x^4 \leq x^2$ when $x \in [-1,1],$ the above inequality may be bound above by
\begin{flalign*}
    0 &\geq -(Cx+Du)^\top \Q (Cx+Du)- 2(Cx+Du)^\top \bS u \\&- u^\top \R u -k_2\bP x^2 + x\bP B u.
\end{flalign*}
Replacing $\Q$, $\bS$, and $\R$ with the associated values in Table \ref{tab:QSR} for non-degenerate interior conic systems and reformulating the above inequality as an \ac{lmi} results in \eqref{eq:conicConditions}. \qed


\end{pf}

\bibliographystyle{plain}        
\bibliography{autosam}           

\end{document}